\documentclass{SciPost}

\hypersetup{
    colorlinks,
    linkcolor={red!50!black},
    citecolor={blue!50!black},
    urlcolor={blue!80!black}
}

\usepackage[bitstream-charter]{mathdesign}
\DeclareSymbolFont{usualmathcal}{OMS}{cmsy}{m}{n}
\DeclareSymbolFontAlphabet{\mathcal}{usualmathcal}
\usepackage[bitstream-charter]{mathdesign}
\usepackage{tcolorbox}
\usepackage{color}
\usepackage{xr}
\usepackage{subcaption}
\usepackage[table]{xcolor}
\usepackage{ragged2e}
\usepackage{verbatim}
\usepackage{fancyvrb}
\usepackage{xcolor}

\definecolor{codebg}{RGB}{248,248,248}
\definecolor{codekw}{RGB}{0,0,180}
\definecolor{codecomment}{RGB}{120,120,120}
\definecolor{codestring}{RGB}{0,120,0}

\newenvironment{codeblock}
  {\VerbatimEnvironment
   \begin{Verbatim}[
     fontsize=\small,
     frame=single,
     rulecolor=\color{black},
     numbers=left,
     numbersep=5pt,
     xleftmargin=1em,
     commandchars=\\\{\},
   ]}
  {\end{Verbatim}}

\definecolor{bg}{gray}{0.95}

\fancypagestyle{SPstyle}{
\fancyhf{}
\lhead{\colorbox{scipostblue}{\bf \color{white} ~SciPost Physics Codebases }}
\rhead{{\bf \color{scipostdeepblue} ~Submission }}

\fancyfoot[C]{\textbf{\thepage}}
}

\begin{document}

\pagestyle{SPstyle}

\begin{center}{\Large \textbf{\color{scipostdeepblue}{
C-Pol: Point charge perturbation scheme for mapping tensor moment surfaces\\
}}}\end{center}

\begin{center}\textbf{
Anoop Ajaya Kumar Nair\textsuperscript{1$\star$},
Julian Bessner\textsuperscript{2}, 
Timo Jacob\textsuperscript{2,3,4} and
Elvar~\"O.~J\'onsson\textsuperscript{1$\dagger$}
}\end{center}

\begin{center}
{\bf 1} Science Institute and Faculty of Physical Sciences, University of Iceland, Reykjav\'ik, Iceland
\\
{\bf 2} Institute of Electrochemistry, Ulm University, Albert-Einstein-Allee 47, 89081 Ulm, Germany
\\
{\bf 3} Helmholtz-Institute Ulm (HIU) Electrochemical Energy Storage, Helmholtzstr. 11, 89081 Ulm, Germany
\\
{\bf 4} Karlsruhe Institute of Technology (KIT), P.O. Box 3640, 76021 Karlsruhe, Germany
\\[\baselineskip]
$\star$ \href{mailto:email1}{\small mailanoopanair@gmail.com}\,,\quad
$\dagger$ \href{mailto:email2}{\small elvarorn@hi.is}
\end{center}

\section*{\color{scipostdeepblue}{Abstract}}
\textbf{\boldmath{
We present an efficient moment-based perturbation scheme for evaluating polarizability tensors of small molecules at a fraction of the computational cost of conventional energy-based approaches. Rather than applying explicit electric fields, the method perturbs the molecular charge density using strategically arranged external point charges, as in QM/MM simulations, and extracts polarizability tensors from finite differences of multipole moments extrapolated to the zero-perturbation limit. C-Pol is implemented as a backend-agnostic Python package requiring only that the host quantum chemistry code support external point-charge potentials. We demonstrate interfaces with three codes spanning complementary numerical approaches: GPAW (real-space grids), NWChem, and PySCF (atom-centered basis sets with complete basis set extrapolation). Validation against Gaussian 16 energy-based reference calculations across molecules representing all 19 commonly occurring point groups yields agreement within 3\%, with the largest deviations confined to tensor components that are numerically small and contribute negligibly to the electrostatic potential. Finite-difference polarizabilities computed via C-Pol are further shown to agree with analytic coupled-perturbed Hartree-Fock values to within 0.02 a.u. across all tested basis sets, at comparable computational cost for medium-sized systems and with more favorable scaling for larger ones.
The package provides a practical and transferable route to high-quality multipole polarizability data for force field development, polarizable embedding, and machine-learning training sets.
}
}

\vspace{\baselineskip}

\noindent\textcolor{white!90!black}{%
\fbox{\parbox{0.975\linewidth}{%
\textcolor{white!40!black}{\begin{tabular}{lr}%
  \begin{minipage}{0.6\textwidth}%
    {\small Copyright attribution to authors. \newline
    This work is a submission to SciPost Physics Codebases. \newline
    License information to appear upon publication. \newline
    Publication information to appear upon publication.}
  \end{minipage} & \begin{minipage}{0.4\textwidth}
    {\small Received Date \newline Accepted Date \newline Published Date}%
  \end{minipage}
\end{tabular}}
}}
}


\vspace{10pt}
\noindent\rule{\textwidth}{1pt}
\tableofcontents
\noindent\rule{\textwidth}{1pt}
\vspace{10pt}


\section{Introduction}
\label{sec:intro}
Molecular moment and polarizability tensors are compressed representations of the charge density distribution of a molecule  \cite{buckingham, applequist,AJStone2008IMP}. Through the use of these parameters, the electrostatic potential landscape is compactly described by a finite expansion; with careful parameterization, this expansion  reproduces the potential generated by the total charge density distribution, typically to within a few percent \cite{Marshall2013}. Consequently, the utilization of moment and polarizability tensors can facilitate an efficient and accurate modeling of intermolecular electrostatic interactions. Conventional techniques for computing molecular polarizability tensors \cite{MAROULIS200816, hohm2006experimental} rely on finite perturbing fields \cite{Cohen1965, MAROULIS1985182,BISHOP1986377}. These techniques mainly  involve either the variation in energy \cite{Loboda2016,CHEN2020137555} (energy-based scheme) or change in the charge density distribution \cite{Elking2011} (density-based/moment-based scheme) of the molecule in the presence of an externally applied potential field or potential field gradient. 

There are numerous examples which demonstrate the calculation of polarizability tensors \cite{Cohen1965,MAROULIS1985182,MAROULIS200816,BISHOP1986377,dykstra1984derivative,Dykstra1985,liu1985polarizabilities,liu1987multipole,Chen_2020}.  Finite field methodologies depending on the energy- or moment-based schemes have been utilized to calculate molecular polarizability tensors, extending up to the quadrupole-quadrupole or dipole-octupole levels, particularly suited for highly symmetrical small molecules or atoms \cite{Cohen1965,MAROULIS1985182,MAROULIS200816,BISHOP1986377}. This avenue of exploration has been further refined by Elking et al. to enable the computation of polarizabilities of arbitrary rank, applicable across molecules of diverse sizes and geometries \cite{Elking2011}. 
These methods mandate that the underlying quantum chemistry code used as the backend calculator should have the feature available to apply a perturbing field and higher order potential-gradients. Such functionality is often implemented in codes that are based on localized atomic orbital basis sets. The implementation of routines to apply potential field gradients, whilst having a field magnitude of zero, is not prevalent in open source codes, and energy-based perturbation schemes require elaborate extrapolation schemes to converge to the basis set limit which are computationally expensive in terms of the number of single point calculations required \cite{Loboda2016}. 

Analytical techniques for polarizability calculations on the other hand, treat the polarizability tensor as an energy derivative with respect to an external electric field and the tensor if obtained by solving iteratively response (perturbation) equations rather than solving finite-difference equations. In the coupled-perturbed Hartree-Fock (CPHF) \cite{cphf1, CPHF1970} or coupled-perturbed Kohn-Sham (CPKS) \cite{CPKS1993} approach, one differentiates the Hartree-Fock/Kohn-Sham equations with respect to the field, leading to linear equations for the first-order change in the molecular orbitals (and hence density), whose solution yields static or frequency-dependent polarizabilities as exact analytic derivatives within the chosen level of theory and basis set \cite{cphf1, cphf2, cphf3}. Compared to finite-field methods, analytic CPHF/CPKS avoids numerical differentiation errors and offers better efficiency and stability, especially when extended to higher-order properties like hyperpolarizabilities \cite{cphf_derv}. A key demerit of analytic CPHF/CPKS polarizability calculations is their computational cost and scaling, especially for large systems, because solving coupled linear-response equations for each field direction and frequency is computationally demanding when compared to just doing a few independent SCF calculations \cite{cphf2, cphf3}. This leads to higher memory demands, more complicated implementations, and potentially poor scaling compared to simple finite-field approaches, which can be a serious limitation for very large molecules or when many frequencies/properties are needed. This is presumably the reason why finite-field energy based methods are comparatively prevalent in literature \cite{Loboda2016}. 

The polarizability tensors find application in the formulation of intermolecular interaction potentials, notably in polarizable force field models. Various polarization models utilize induced atomic or molecular dipoles such as in the Drude oscillator model \cite{lamoureux2003simple}, Thole Type Models (TTM) \cite{ttm2, ttm3, ttm4}, AMOEBA \cite{amoeba}, HIPPO \cite{HIPPO}, MB-Pol \cite{mbpol} or OCP3-Pol \cite{opc3pol}, and even higher order polarizabilities such as SCME \cite{Wikfield2013,EOJ2022,Myneni2022}, which includes the dipole-quadrupole and quadrupol-quadrupole polarizability. 

In addition, recent neural-network and kernel-based models that explicitly target molecular response properties such as dipole moments and polarizability tensors from three-dimensional geometries have created a clear demand for efficient, scalable workflows to generate large, high-quality training datasets. Symmetry-adapted Gaussian process approaches like AlphaML demonstrate that coupled-cluster-level polarizabilities can be accurately learned at negligible marginal cost relative to electronic-structure calculations \cite{alphaML}, while $\Delta$-machine-learning schemes that correct lower-level polarizable dipole models toward high-level references further emphasize the importance of rich, systematically improvable datasets \cite{DeltaML}. At the same time, advances in rotation‑equivariant graph neural networks, such as MGNN for universal molecular potentials \cite{MGNN}, together with symmetry‑adapted models for coupled-cluster polarizabilities and high‑order equivariant message-passing architectures like MACE‑MDP for accurate prediction of IR and Raman spectra using machine-learned dipoles and dipole-dipole polarizabilities\cite{MACEMDP}, underscore the growing role of machine-learning frameworks that can simultaneously capture energies, forces, and response properties including dipole moments and polarizability tensors. Hence highlighting the opportunity to extend these ideas to higher multipole moments and thus motivate automated pipelines for generating comprehensive multipole and polarizability tensor data across chemically diverse compound spaces.

Even though the importance of calculating moments and polarizabilities has been established 
the procedure for calculating the polarizability tensors is not standardized. Most quantum chemistry codes, apart from a select few, are only capable of calculating the dipole-dipole polarizability ($\alpha$). The few that are able to output higher order polarizabilities do so with certain set-backs. Higher order polarizabilities like the dipole-quadrupole ($A$) and quadrupole-quadrupole ($C$) polarizability tensors can be obtained in the traceless form from CADPAC \cite{amos1987cadpac}. MOLPRO \cite{Molpro} and DALTON \cite{dalton} 
are capable of calculating traced cartesian molecular polarizability tensors utilizing a range of quantum chemistry methods, including standard SCF (such as DFT and HF) and electron correlation methods such as MP2 and CCSD \cite{Elking2011}. 
This implies that the range of approximate methods and basis sets available for calculating polarizability tensors is constrained by the features of the aforementioned quantum chemistry codes. 

To rectify issues associated with established implementations we propose a generalized charge-based perturbation scheme which can be easily coupled to existing and open source quantum chemistry (QC) codes as a light weight python wrapper. This is achieved by introducing perturbing external potentials which results from simple arrangements of point charges, such as are routinely applied in hybrid quantum mechanics / molecular mechanics (QM/MM) simulation methods. 

We introduce a modular Python package called C-Pol and demonstrate its application to several quantum chemistry codes. Specifically, we apply it to the open-source, grid-based GPAW \cite{gpaw, gpaw2, gpaw3} code which includes a built-in QM/MM interface \cite{QMMM1} to compute the $\alpha$, $A$, and $C$ polarizability tensors directly on the grid, thereby eliminating the need for basis set extrapolation. We further extend C-Pol to NWChem \cite{nwchem}, and PySCF \cite{pyscf} by implementing it as an external wrapper. Both NWChem and PySCF are used to compute multipole polarizability tensors across a hierarchy of correlation-consistent basis sets (pVDZ, pVTZ, and pVQZ), with the results extrapolated to the complete basis set (CBS) limit \cite{CBS}. In our method, the molecule in the QM region is perturbed by fields and field-gradients via a set of point charges positioned in such a way as to produce exclusively only fields or only field-gradients at the chosen center of the expansion of the concerned charge distribution. 

The strategic distribution of the point charges to produce field and field gradient tensors with predictable geometry and magnitude in combination with finite-difference grid-based description of wavefunctions enables a reduction in the number of SCF computations by a factor of $\sim$30 compared to conventional and established energy-based schemes which require complete basis set extrapolation.
The C-Pol package is designed with a backend-agnostic architecture, allowing polarizability tensor calculations to be driven by any QM code with a QM/MM method implementation that can accept external point-charge perturbations and return the corresponding multipole moments. The package exposes a unified Python interface centred on the \texttt{Pols} class that accepts an \textsc{ase} \texttt{Atoms} \cite{ASE1} object as input (as a geometry input), a \texttt{code} variable, which indicated the desired QM calculator to be used as well as the parameters associated with it (via the \texttt{params}).   and the methods \texttt{calc\_dd\_pol()}, \texttt{calc\_dq\_pol()}, and \texttt{calc\_qq\_pol()}, internally constructs and performs the field/field-gradient perturbed QM calculations. Examples of backend incorporation of QM codes have been explored using  GPAW, NWChem and PySCF and has been extensively documented \cite{cpol_docs}.

\section{Methods}
\label{section:theory}

The electrostatic energy of a molecule can be expanded as a Taylor series in terms of molecular moment tensors (monopole, $q$, dipole, $\mu$, quadrupole, $\theta$, octupole, $\Omega$...), and molecular polarizabilities (dipole-dipole, $\alpha$, dipole-quadrupole, $A$, quadrupole-quadrupole, $C$, dipole-octapole, $D$... ) and the hyperpolarizabilities (first, $\beta$, second $\gamma$...) in response to external perturbation in the form of a potential, $V(\mathbf{r})$,
resulting in the Buckingham expansion of the potential energy \cite{buckingham}. In terms of traceless moments it reads
\begin{align}
    U[V(\mathbf{r})] &= U_o + qV +\mu^{o}_{\alpha}V_{\alpha} -\frac{1}{2}\alpha_{\alpha\beta}V_{\alpha}V_{\beta} \nonumber \\
     &-\frac{1}{6}\beta_{\alpha\beta\gamma}V_{\alpha}V_{\beta}V_{\gamma} +\frac{1}{3}\theta^{o}_{\alpha\beta}V_{\alpha\beta} \nonumber \\
     &-\frac{1}{3}A_{\gamma,\alpha\beta}V_{\gamma}V_{\alpha\beta} - \frac{1}{6}C_{\alpha\beta,\gamma\delta}V_{\alpha\beta}V_{\gamma\delta} \nonumber \\
     &-\frac{1}{15}\Omega^{o}_{\alpha\beta\gamma}V_{\alpha\beta\gamma} - \frac{1}{15}D_{\delta,\alpha\beta\gamma}V_{\delta}V_{\alpha\beta\gamma} + \dots \label{bucking_eq}
\end{align}
Here $U_o$ is the internal energy of the unperturbed system, $V$ is the Coulomb potential at a point representative of the molecule (or atom), most often chosen as the center of mass \cite{batista1998molecular}, and the potential field and higher order fields are derived by repeated application of the gradient operator on $V$, and is given by $V_\alpha = \nabla_\alpha V, V_{\alpha\beta} = \nabla_\beta V_\alpha\dots $. It can be observed that each additional multipole term of rank 'n' adds a contribution that depends on $1/r^{(2n+1)}$ to the electrostatic energy, where $r$ is the distance between the source and site. Hence, the series is often terminated at the term representing the energy due to the hexadecapole moment or lower 
since the contribution of higher order multipole ranks is irrelevant to the overall intermolecular interaction energy \cite{Batista_MMMWMIh}. Most polarizable force-fields potential functions tend to terminate at lower terms for computational efficiency \cite{lamoureux2003simple}.

\subsection{Calculating the polarizability tensors}

With the expansion of the electrostatic energy it is trivial to derive the polarizability tensors in terms of potential fields and potential field gradients. For example, 
taking derivative of equation \ref{bucking_eq} with respect to the potential field ($V_{\sigma}$) results in equation \ref{eq:dudv_sig}.
\begin{align}
    \frac{\partial U}{\partial V_{\sigma}} &= \mu^{o}_{\sigma} -\alpha_{\sigma\beta}V_{\beta} -\frac{1}{2}\beta_{\sigma\beta\gamma}V_{\beta}V_{\gamma} \nonumber \\
     &-\frac{1}{3}A_{\sigma,\alpha\beta}V_{\alpha\beta} - \frac{1}{15}D_{\sigma,\alpha\beta\gamma}V_{\alpha\beta\gamma} + \dots
     \label{eq:dudv_sig}
\end{align}
For a given perturbing potential field and potential field gradient we can equate the change in electrostatic energy to the change in the dipole moment as 
\begin{align}
    \mu'_\sigma \equiv \frac{\partial U}{\partial V_{\sigma}} &= \mu^{o}_{\sigma} - \alpha_{\sigma\beta}V_{\beta} - \frac{1}{2}\beta_{\sigma\beta\gamma}V_{\beta}V_{\gamma} \nonumber \\
     &-\frac{1}{3}A_{\sigma,\alpha\beta}V_{\alpha\beta} - \frac{1}{15}D_{\sigma,\alpha\beta\gamma}V_{\alpha\beta\gamma} + \dots
     \label{eq:mu_full}
\end{align}
Given a small perturbing field (and a zero field gradient), and ignoring (due to insignificant contribution) the terms containing field squared ($V_{\alpha}V_{\beta}$) and higher order terms and field gradient components of rank 3 ($V_{\alpha \beta\gamma}$) and greater ,
we can isolate the dipole-dipole polarizability ($\alpha$) as: 
\begin{align}
\alpha_{\sigma\beta} =& -\frac{\mu'_\sigma - \mu_\sigma}{V_\beta}
\label{eq:dip-dip}
\end{align}

Similarly, if we set the potential field ($V_{\gamma},V_{\beta}$) and rank-3 or higher order terms equal to zero and maintain a non-zero potential field gradient ($V_{\alpha\beta}$) then the dipole-quadrupole polarizability ($A$)  is formulated as:
\begin{align}
    A_{\sigma,\alpha\beta} &= -3\frac{\mu^{'}_{\sigma}-\mu^{o}_{\sigma}}{V_{\alpha\beta}}
    \label{eq:dip-quad}
\end{align}

In order to calculate the quadrupole-quadrupole polarizability ($C$), the derivative of the electrostatic energy w.r.t the field-gradient vector ($V_{\sigma\eta}$) as mentioned by Buckingham et.al \cite{buckingham}: 

\begin{align}
    \theta'_{\sigma\eta} \equiv -3\frac{\partial U}{\partial V_{\sigma\eta}} & =  \theta^{o}_{\sigma\eta} - A_{\gamma,\sigma\eta}V_{\gamma} -C_{\alpha\beta,\sigma\eta}V_{\alpha\beta} + \dots
\end{align}
and as in the calculation of $A_{\sigma, \alpha\beta}$ if we set all other gradient ranks to zero we get:
\begin{align}
    C^{o}_{\alpha \beta,\sigma\eta} &= -\frac{\theta^{'}_{\sigma\eta}-\theta^{o}_{\sigma\eta}}{V_{\alpha\beta}}
    \label{eq:quad-quad}
\end{align}
The number of non-zero unique components in the polarizability tensors is related to the point group of the molecule under consideration \cite{buckingham}. In order to achieve the irreducible representation of the polarizability tensors, the molecule under consideration must be oriented with respect to the x, y and z axes, so as to maximize the number of symmetry elements along each of these axes. 

\noindent

\begin{figure}[!th]
     \begin{center}
     \includegraphics[width=0.5\paperwidth]{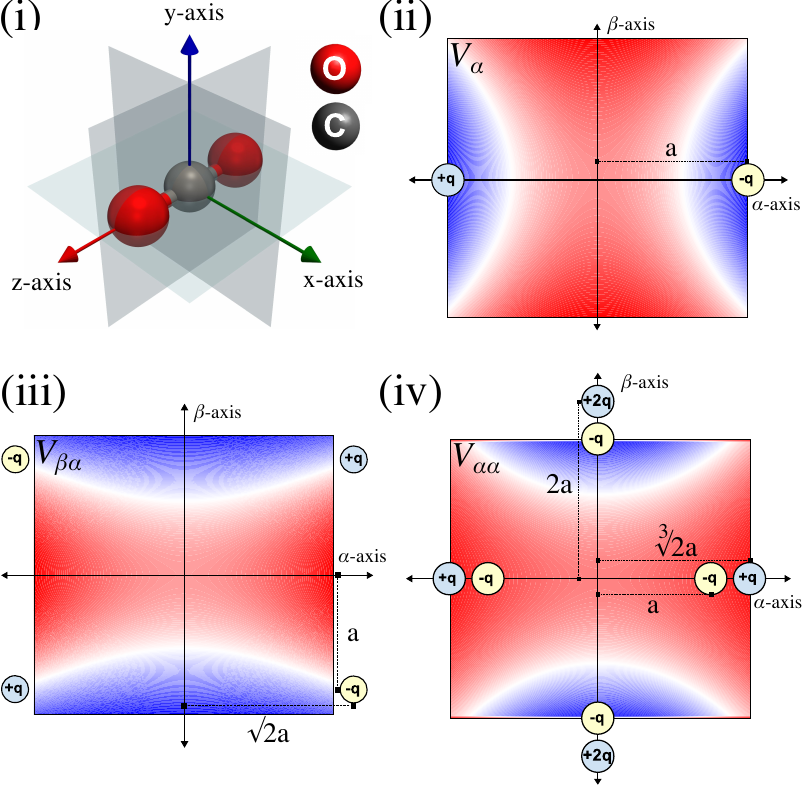}
     \end{center}
     \caption{\textbf{Charge perturbation scheme:} (i) An example molecular system, CO$_2$, where the principal axis ($z$-axis) is aligned along the symmetry axis. This is the standard orientation of the CO$_2$ molecule that maximizes the number of symmetry elements along each axis (x, y and z axis). (ii) Point charge distribution for creating a potential field (e/$a_{0}^2$). (iii) and (iv) point charge distributions for creating non-zero $V_{\alpha\beta}$ and $V_{\alpha\alpha}$ potential field gradient (e/$a_{0}^3$), respectively.}
     \label{fig:cd}
\end{figure}

This is indicative of the so called "standard orientation" presented in the Gaussian 16 documentation \cite{g16}. The center of mass of the molecule should also coincide with the origin of the coordinate system. The standard orientation of the $CO_2$ molecule is presented in Figure \ref{fig:cd}(i). The orientation of molecules possessing varied point group symmetries are provided in Section 6 of the Supplementary Information (SI).

\subsection{Charge based density perturbation}

To generate the three components of the potential field vector (namely $V_x$, $V_y$ and $V_z$) and the six independent components of the potential field gradient tensor (namely $V_{xx}$, $V_{yy}$, $V_{zz}$, $V_{xy}$, $V_{xz}$ and $V_{yz}$), we can use the point charge distributions given in Figure \ref{fig:cd}(ii)-(iv). 
Here $\alpha$, $\beta$, $\gamma$ can acquire a certain value from the set of \{x,y,z\} axes.
Each distribution is characterized by a given value of the distance parameter $\textbf{a}$ (since the distances of the MM point charges from the origin are defined as $\textbf{a}$ multiplied with a pre-factor), and charge magnitude $\textbf{q}$. The magnitude of the perturbations in terms of $\textbf{a}$ and $\textbf{q}$ are given in the equations below. In order to obtain a charge distribution which nets a certain magnitude of the field, or field gradient, we first fix the magnitude of $\textbf{a}$ (distance unit) and scale the magnitude of the charge $\textbf{q}$.  

\begin{equation}
    V_{i} = \frac{-2q}{a^2} \implies q = \frac{-V_{i}a^2}{2}
\end{equation}

\begin{equation}
    V_{ii} = \frac{-3q}{2a^3} \implies q = \frac{-2V_{ii}a^3}{3}
\end{equation}

\begin{equation}
    V_{ij} = \frac{4\sqrt{2}q}{(\sqrt{3}a)^3} \implies q = \frac{(\sqrt{3}a)^3V_{ij} }{4\sqrt{2}}
\end{equation}

The derivation of the field and field-gradient tensor components is elaborated in Section 2 of the SI. 
For example, the potential field component $V_{\alpha}$ (of magnitude $|V_\alpha|$) is produced at the center of the distribution scheme in Figure~\ref{fig:cd}(ii).
A potential field gradient with a non-zero off-diagonal tensor component of magnitude $|V_{\beta\alpha}|$ is obtained using the charge distribution scheme shown in Figure~\ref{fig:cd}(iii). Similarly, to produce a potential field-gradient tensor $V_{\alpha\alpha}$ with magnitude $|V_{\alpha\alpha}|$, and $-|V_{\alpha\alpha}|$ for the $V_{\gamma\gamma}$ component, the charge distribution scheme in Figure~\ref{fig:cd}(iv) is used. The appearance of two components, $V_{\alpha\alpha}$ and $V_{\gamma\gamma}$, with equal and opposite magnitudes $|V_{\alpha\alpha}|$ and $-|V_{\alpha\alpha}|$ respectively, follows directly from Laplace's equation, which requires the electric field gradient tensor to be traceless, i.e.\ $\sum_{\alpha} V_{\alpha\alpha} = 0$.

While this work focuses on the $\alpha$, $A$ and $C$ polarizability tensors the charge perturbation scheme can be generalized to higher order perturbations, which is outlined in Section 3 of the SI.

\subsection{Extrapolation scheme}
\label{method:ex_scheme}

The potential field and potential field-gradient components acquire the values mentioned previously at the center of symmetry of the charge distribution. 
Ideally, a constant value is preferred throughout the grid. but this is not possible with the charge distributions given in Figure~\ref{fig:cd}(ii)-(iv) and there are non-zero potential field and potential field-gradient components present in regions around the center (of the charge distribution) where they should ideally be absent.  

\begin{figure*}[!ht]
     \begin{center}
     \includegraphics[width=0.7\paperwidth]{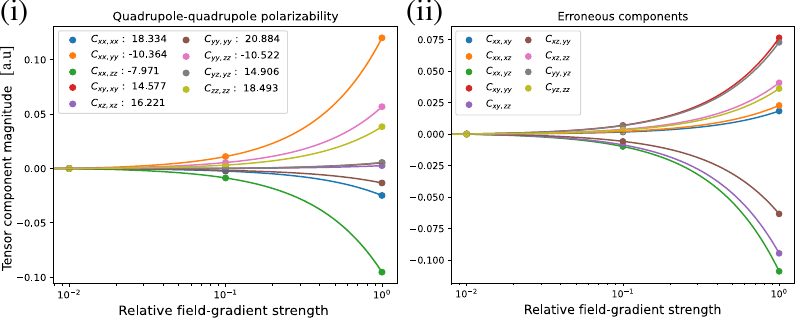}
     \end{center}
     \caption{The extrapolations scheme applied to the calculated tensor moments of a H$_2$O molecule. (i) Shows the active quadrupole-quadrupole tensor components as dictated by the symmetry of the molecule, which converge to a constant value. The constant value is given in the legend. (ii) In-active quadrupole-quadrupole components which converge to zero at the zero perturbation limit.
     The tensor components are base-line shifted with respect to the extrapolated value. All the C tensor components are presented in atomic units ($a_0^5$)}
    \label{fig:extrapolate}
\end{figure*}

However, the error introduced by these non-zero components can be eliminated by extrapolating to the infinitesimal perturbation limit, as illustrated below by considering the $\alpha$ polarizability tensor as an example.
In equation \ref{eq:mu_full}, if  $-\delta\mu'/\delta V_{\eta}$ is calculated without truncating \ref{eq:mu_full} at $V^2$, we get equation \ref{eq:alpha_full}. 

\begin{align}
    \frac{-\delta(\mu'_\sigma - \mu^{o}_{\sigma})}{\delta V_{\eta}} \equiv \alpha'_{\sigma\eta} &=   \alpha^o_{\sigma\eta} + \beta_{\sigma\eta\gamma}V_{\gamma} + O(V^2)\label{eq:alpha_full}
\end{align}

From equation \ref{eq:alpha_full} it is clear that the dipole-dipole polarizability ($\alpha'$) converges to a constant value ($\alpha^{o}$) as the perturbative field is infinitesimal ($V_{\gamma} \to 0$). This is what we refer to as a "zero-perturbation limit". 

To achieve this, three perturbative calculations are performed, with the magnitude of the applied field and field-gradient reduced by a factor of 10 in each successive calculation. The zero-perturbation limit of each tensor component is then obtained by linear extrapolation over these three values. Tensor components that are required to vanish by the point group symmetry of the molecule ($\alpha^{'}$) extrapolate to zero, while non-zero components converge to their corresponding physical values ($\alpha^{o}$).
A similar linear extrapolation approach can be taken to converge the $A$ and $C$ polarizability tensors.

This extrapolation is illustrated in Figure \ref{fig:extrapolate}(i) and (ii) for the quadrupole-quadrupole polarizability of the H$_2$O molecule.

\section{Code}
\label{method:implementation}

To enable flexible and backend-independent evaluation of multipole moments and polarizability tensors, \texttt{C-Pol} is structured as a lightweight python layer interfacing with different electronic-structure codes. A routine to obtain moments (here, dipoles and quadrupoles) is implemented for each backend QM-code, that evaluates the perturbed moment for a molecule under the influence of external point charges. A separate class (\texttt{Pols}) then constructs the polarizability tensor by finite differences. The signature for the moment calculation function that acquires the multipole moments under the influence of a field or field gradient created by point charges is as follows:

\begin{codeblock}
\textcolor{codekw}{def} multipole_func(molecule, q_position, q_magnitude, qm_parameters):

    \textcolor{codestring}{"""Return the perturbed dipole/quadrupole (multipole) for a QM}
    \textcolor{codestring}{system in external point charges."""}
    \textcolor{codekw}{return} multipole
\end{codeblock}

The function arguments and the return quantity  of the \texttt{multipole\_func} function is described in table \ref{tab:moment_func_args}.
\begin{table}[!ht]
    \centering
    \rowcolors{2}{gray!25}{white}
    \renewcommand{\arraystretch}{1.5}
    \begin{tabular}{|p{0.25\linewidth}|p{0.25\linewidth}|p{0.38\linewidth}|}
        \hline
        \textbf{Quantity} & \textbf{Type} & \textbf{Description} \\
        \hline
        \texttt{molecule} & ASE \texttt{Atoms} & Molecule structure present in the QM region. \\
        \texttt{q\_position} & array-like $(N,3)$ & Cartesian positions of the external point charges. \\
        \texttt{q\_magnitude} & array-like $(N)$ & Magnitudes of the external point charges. \\
        \texttt{qm\_parameters} & \texttt{dict} & Arguments specific to the back-end QM code for  specifying basis-sets, xc functional.. etc. \\
        \texttt{multipole} & float / \texttt{ndarray} & Perturbed multipole moment tensor. \\
        \hline
    \end{tabular}
    \caption{Arguments and return value of the backend \texttt{multipole\_func} routine.}
    \label{tab:moment_func_args}
\end{table}

The \texttt{multipole\_func} function is invoked to compute the molecular polarizability tensor at a user-specified perturbation strength, by applying either a homogeneous electric field or an electric-field gradient, depending on the rank of the polarizability tensor of interest ($\alpha$, $A$ and $C$). The function signature is provided below:

\begin{codeblock}
\textcolor{codekw}{def} polarizability_func(molecule, perturbation_strength, qm_parameters):

    \textcolor{codestring}{"""Assemble the finite-difference polarizability tensor from}
    \textcolor{codestring}{perturbed multipoles."""}
    \textcolor{codekw}{return} polarizability
\end{codeblock}

The function arguments and the return quantity of the \texttt{polarizability\_func} function is described in table \ref{tab:polarizability_args}.

\begin{table}[!ht]
    \centering
    \rowcolors{2}{gray!25}{white}
    \renewcommand{\arraystretch}{1.5}
    \begin{tabular}{|p{0.25\linewidth}|p{0.25\linewidth}|p{0.38\linewidth}|}
        \hline
        \textbf{Quantity} & \textbf{Type} & \textbf{Description} \\
        \hline
        \texttt{molecule} & ASE \texttt{Atoms} & Molecular structure of the quantum-mechanical region. \\
        \texttt{perturbation strength} & float or array-like & Magnitude of the applied external field/field-gradient. \\
        \texttt{qm\_parameters} & \texttt{dict} & Arguments specific to the back-end QM code for  specifying basis-sets, xc functional.. etc. \\
        \texttt{polarizability} & \texttt{ndarray} & Polarizability tensor assembled from perturbed multipoles. \\
        \hline
    \end{tabular}
    \caption{Arguments and return value of the \texttt{polarizability\_func} calculation routine.}
    \label{tab:polarizability_args}
\end{table}

The following three examples illustrate this strategy in progressively more technical detail: (i) a direct source-level integration into GPAW, where the all-electron density on the real-space grid is used to evaluate multipole moments and polarizabilities; (ii) a wrapper around NWChem that generates, runs, and parses DFT input/output files; and (iii) a PySCF-based implementation that works directly with the real-space electron density on a numerical grid (cube-format).

\subsection{Source code modification: GPAW as an example}

The \texttt{C-Pol} code has been integrated into GPAW by adding a dedicated C module (\texttt{moments.c}) that computes multipole moments from the dipole up to the hexadecapole. These moments are obtained from the charge density interpolated to a fine real-space grid mesh, accessed via the \texttt{calc.density.finegd} attribute. External electric fields and their spatial gradients are modeled by point charges introduced as molecular mechanics (MM) sites using the \texttt{PointChargePotential} class defined in \texttt{gpaw/external.py}. The moments (dipole to hexa-decapole) under zero perturbation are accessed within the \texttt{MomTensor} class. On the other hand, moments under different field and field-gradient perturbation tensors are used to construct the different polarizability tensors ($\alpha$, $A$, and $C$) within the \texttt{PolarTensor} class. To run an example calculation, after cloning the GPAW C-Pol repository \cite{gpaw_cpol} and installing it, the example calculation can be run via the example given below:

\begin{codeblock}
\textcolor{codekw}{from} ase.io \textcolor{codekw}{import} read
\textcolor{codekw}{from} gpaw \textcolor{codekw}{import} GPAW
\textcolor{codekw}{from} gpaw.eigensolvers \textcolor{codekw}{import} RMMDIIS
\textcolor{codekw}{from} gpaw.multipoles \textcolor{codekw}{import} MomTensor, PolarTensor

\textcolor{codecomment}{# Read geometry from XYZ}
atoms = read(\textcolor{codestring}{"h2o.xyz"})  \textcolor{codecomment}{# assumes H2O in this file}
atoms.center(vacuum=3.5)

\textcolor{codecomment}{# GPAW calculator}
atoms.calc = GPAW(mode=\textcolor{codestring}{"fd"}, 
                  h=0.20, 
                  xc=\textcolor{codestring}{"PBE"},
                  convergence=\{\textcolor{codestring}{"eigenstates"}: 1e-6,
                               \textcolor{codestring}{"density"}:1e-8\},
                  eigensolver=RMMDIIS(niter=5))

\textcolor{codecomment}{# Multipole moments}
moments = MomTensor(atoms).get_multipole_moments()

\textcolor{codecomment}{# Polar tensors}
pol = PolarTensor(atoms)
dd = pol.calc_dd_pol()
dq = pol.calc_dq_pol()
qq = pol.calc_qq_pol()
\end{codeblock}

A detailed description of the implementation and its usage is provided in the \texttt{C-Pol} documentation \cite{cpol_docs}.

\subsection{Developing a wrapper: NWChem as an example}
Polarizability tensors are computed using \texttt{C-Pol}, a Python wrapper interfacing with the NWChem quantum chemistry package. The perturbed/unperturbed dipole and quadrupole moments are extracted by parsing the NWChem output files directly. To obtain the field-perturbed counterparts, the \texttt{Dipole} and \texttt{Quadrupole} wrapper functions accept an ASE \texttt{Atoms} object together with a set of external point charge positions and magnitudes, which constitute the molecular mechanics (MM) point charge perturbation. These point charges impose either a uniform electric field $V_{\alpha}$ or an electric field gradient $V_{\alpha\beta}$ on the quantum mechanical (QM) molecule, thereby simulating the desired perturbation environment. For each perturbation scenario, the wrapper (via the \texttt{create\_nw\_input function}) constructs a NWChem DFT input file, embeds the point charges via the \texttt{bq} block, executes the calculation (via a python \texttt{subprocess}), and parses the resulting output (using the \texttt{extract\_moment} function) to extract the perturbed dipole moment $\mu'$ or quadrupole tensor $\Theta'$. An example of the function signature for the \texttt{Dipole} wrapper, implementing the procedure described above, is provided below. These perturbed quantities are subsequently passed to the \texttt{Pols} class, which assembles the complete set of molecular polarizability tensors (Eqs. \eqref{eq:dip-dip}–\eqref{eq:quad-quad}).

\begin{codeblock}
\textcolor{codekw}{def} Dipole( molecule, 
            charge_postions, 
            charge_mag, 
            params):
    SCRATCH_DIR = get_scratch_dir()
    input_file = create_nw_input(molecule, 
                                 charge_postions, 
                                 charge_mag,
                                 params, 
                                 \textcolor{codestring}{"DIPOLE"}, 
                                 ...)
    with open(os.path.join(SCRATCH_DIR, \textcolor{codestring}{"prefix.nwi"}), \textcolor{codestring}{'w'}) as f:
        f.write(input_file)
    subprocess.run(f\textcolor{codestring}{"nwchem prefix.nwi > prefix.nwo"}, shell=True)
    \textcolor{codekw}{return} extract_dpole(os.path.join(SCRATCH_DIR, "prefix.nwo"))
\end{codeblock}

The \texttt{Quadrupole} function follows the same pattern, requesting the \texttt{QUADRUPOLE} property and returns the symmetric 3×3 tensor assembled from the six independent components ($Q_{xx}$, $Q_{yy}$, $Q_{zz}$, $Q_{xy}$, $Q_{xz}$, and $Q_{yz}$) parsed from the NWChem output. The C-Pol documentation outlines the NWChem implementation extensively \cite{nwchem_cpol}. A similar wrapper has also been developed for the Psi4 code\cite{psi4} ,
enabling a direct comparison between the moment-based approach used in the C-Pol
implementation and the analytical computation of the dipole-dipole
polarizability ($\alpha$) via the coupled-perturbed Hartree-Fock (CPHF) method (as implemented in Psi4).

\subsection{Utilizing electron density in cube format: PySCF as an example}

The PySCF backend employs an analogous wrapper interface; however, it
evaluates multipole moments by projecting the electron density onto a
real-space grid following the cube-file convention (a procedure adopted
specifically for the electronic quadrupole moment calculation). 

For the perturbed dipole moments, the \texttt{Dipole} function handles
QM/MM embedding through \texttt{pyscf.qmmm.mm\_charge}, which folds the
external point charges directly into the restricted Kohn-Sham (RKS)
Hamiltonian; the resulting dipole moment is then retrieved via   
\texttt{mf.dip\_moment()}. The \texttt{Quadrupole} function introduces an additional post-SCF step: once the perturbed RKS wavefunction has converged, the real-space
electron density ($\rho$) is evaluated on a uniform cubic grid using
\texttt{pyscf.dft.numint}, and the electronic contribution to the
quadrupole moment is obtained by numerical integration of the traceless
quadrupole kernel defined in Eq.~\eqref{eq:qpole_kernel}.

\begin{equation}
    Q_{\alpha\beta} = \frac{3}{2}
    \int \rho(\mathbf{r})
    \left( r_\alpha r_\beta - \frac{1}{3} r^2 \delta_{\alpha\beta} \right) dV.
    \label{eq:qpole_kernel}
\end{equation}
The nuclear contribution is added analytically using Eq. \eqref{e:quad_nuc}.

\begin{align}\label{e:quad_nuc}
    \theta^{nuclear}_{\alpha\beta} = \Sigma_{a} Q^{N}_{a}(\frac{3}{2}a_{\alpha}a_{\beta} - \frac{1}{2}a^{2}\delta_{\alpha\beta})
\end{align}

The quantities $Q^{N}$ and $a$ represent the nuclear point charges (or atomic numbers) and the distance vector. 

\begin{codeblock}
\textcolor{codekw}{def} Quadrupole(molecule, charge_postions, charge_mag, params):
    mol = gto.Mole()
    mol.unit = params[\textcolor{codestring}{"unit"}]
    mol.basis = params[\textcolor{codestring}{"basis"}]
    mol.atom = [[sym, tuple(r)] 
                for sym, r in
                zip(molecule.get_chemical_symbols(), 
                molecule.get_positions())]
    mf = qmmm.mm_charge(params[\textcolor{codestring}{"method"}](mol), 
                                charge_postions, 
                                charge_mag, 
                                unit = params[\textcolor{codestring}{"unit"}])
    mf.xc = params[\textcolor{codestring}{\textcolor{codestring}{"xc"}}]
    mf.kernel()
    rho, n, h_v, coordMod = density(mol, mf.make_rdm1())
    Q_elec = electronic_quadrupole_moment (h_v, 
                                           calcCOM(mol), 
                                           rho, n, 
                                           coordMod)
    Q_nuc  = nuclear_quadrupole_moment(mol)
    \textcolor{codekw}{return} np.round(Q_nuc - Q_elec, 4)
\end{codeblock}

A example of using \textbf{pyscf} to calculate the multipole polarizabilities are provided below:

\begin{codeblock}
\textcolor{codekw}{from} pathlib \textcolor{codekw}{import} Path
\textcolor{codekw}{from} ase.io \textcolor{codekw}{import} read
\textcolor{codekw}{from} pyscf \textcolor{codekw}{import} scf
\textcolor{codekw}{from} cpol_pyscf.poltensor \textcolor{codekw}{import} Pols

atoms = read(Path(__file__).parent / \textcolor{codestring}{"molecule.xyz"})
pols = Pols(atoms, 
            params=\{\textcolor{codestring}{"basis"}: \textcolor{codestring}{"aug-cc-pvtz"},
                    \textcolor{codestring}{"xc"}: \textcolor{codestring}{"pbe"},
                    \textcolor{codestring}{"method"}: scf.RKS,
                    \textcolor{codestring}{"unit"}: \textcolor{codestring}{"au"}\}, 
            code=\textcolor{codestring}{"pyscf"})
            
pols.calc_dd_pol()
pols.calc_dq_pol()
pols.calc_qq_pol()
\end{codeblock}

The C-Pol documentation explains extensively the implementation method \cite{pyscf_cpol}.

\section{Computational details}
In GPAW, the pseudo-wavefunctions are represented on a finite-difference real-space grid with a spacing of 0.18~\AA, and the eigenstates and electron
density are converged to a tolerance of $10^{-8}$ a.u. The RMMDIIS eigensolver is employed for all calculations. All molecular geometries are
first optimized using the Broyden-Fletcher-Goldfarb-Shanno (BFGS) algorithm, with a maximum force component convergence criterion of $5\times10^{-3}$ ~eV/\AA, as implemented in the Atomic Simulation
Environment package~\cite{ASE1}. The Perdew-Burke-Ernzerhof (PBE) \cite{Perdew1996,Perdew1997} exchange-correlation functional is used for all GPAW calculations, except for the validation runs against
Gaussian~16 (G16), where the Becke-Lee-Yang-Parr (BLYP) \cite{B88,LYP} functional is additionally employed. The polarizabilities are calculated using an extrapolation scheme (see section \ref{method:ex_scheme}) with a field (and field gradient) strength of 0.009 a.u (0.0003 a.u), 0.0009 a.u (0.00003 a.u) and 0.00009 a.u (0.000003 a.u), so as to make sure the tensor components converge to the zero perturbation asymptote. The dipole and quadrupole moment integrals are in the traceless form (see Appendix \ref{app:GPAW}), so no additional modification is required. The values for grid spacing, vacuum and field/field-gradient strengths were chosen after extensive convergence tests on a H$_2$O$_2$ system (see Results and Section 5 in the SI). 

For the method validation presented in the Results section, G16 is used to compute the polarizability tensors $\alpha$, $A$, and $C$. These calculations are performed with the PBE functional and, in analogy to the GPAW setup, repeated with BLYP for comparison. In all calculations symmetry detection was turned-off (to prevent it from influencing the symmetry of the polarizability tensors)  and an SCF convergence tolerance of the wavefunctions is set to $10^{-10}$ a.u..  The CBS limit for the energies in each perturbation scenario is extrapolated from energies calculated using the aug-cc-pVDZ, aug-cc-pVTZ and aug-cc-pVQZ basis sets (see Appendix \ref{app:CBS}) to eliminate basis-set dependence. All molecules are relaxed using the corresponding basis-set, prior to application of the energy-based scheme. The polarizabilities (specifically A and C) are de-traced following the method outlined in the Section 1 of the SI. For the NWChem and PySCF calculations, the self-consistent field (SCF) procedure is carried out within a restricted Kohn-Sham (RKS) formalism
using the PBE exchange-correlation functional, and the resulting properties are likewise extrapolated to the CBS limit. The convergence tests with respect to field/field-gradient perturbation and basis-set size are presented in section 5 of the SI.

In order to benchmark the moment-based implementation against an analytical
response method such as coupled-perturbed Hartree-Fock (CPHF), we performed
reference calculations using a C-Pol wrapper around Psi4 to compute the
dipole-dipole polarizability $\alpha$ via the QMMM interface, alongside the
native CPHF response formalism for the water monomer. For each approach,
static Cartesian polarizability tensor components $\alpha_{x,x}$,
$\alpha_{y,y}$, and $\alpha_{z,z}$ were evaluated at the Hartree-Fock level
employing Dunning's augmented correlation-consistent basis sets
(aug-cc-pVDZ, aug-cc-pVTZ, and aug-cc-pVQZ), followed by a simple
complete-basis-set (CBS) extrapolation (as in the case of NWChem and PySCF). The QMMM and CPHF calculated $\alpha$ tensor elements and wall clock timings are reported in Table~\ref{Tab:h2ocphf}. To investigate emergent properties such as the electrostatic potential when expressed as a function of the calculated multipole moments and polarizabilities, a direct comparison with the QM electrostatic potential computed in GPAW was performed. This was done using multipoles obtained following the self-consistent field (SCF) procedure described in Appendix \ref{app:SCF}.

\begin{figure*}[!th]
    \centering
    \includegraphics[width=0.7\paperwidth]{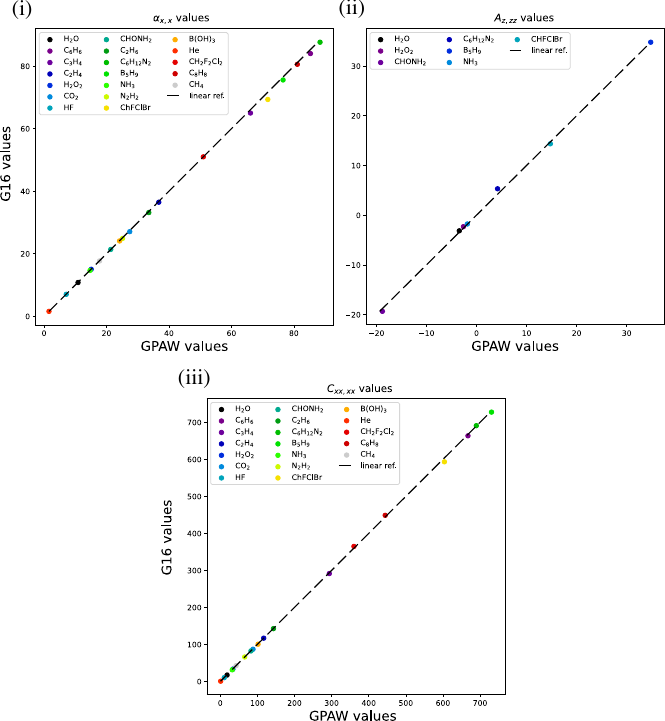} 
    \caption{\justifying{Comparison between the polarizabilities calculated with the moment- and energy-based schemes. The polarizability components for a few example molecules, calculated via energy-based scheme (G16 : y-axis) is plotted against the moments calculated with the moment-based scheme (GPAW: x-axis). (i) The $\alpha_{x,x}$ component ($a_{0}^3$). (ii) The $A_{z,zz}$ component ($a_{0}^4$). (iii) The $C_{xx,xx}$ component ($a_{0}^5$).}} 
    \label{fig:compPol}
\end{figure*}

\section{Results}

\subsection{Validation}

The moment-based polarizability tensors (C-Pol implementation in GPAW) were validated by comparing them, both qualitatively and quantitatively, to tensors obtained from the energy-based scheme (using in Gaussian~16 (G16)) for a set of molecules from different point groups. We verified the implementation at two levels: (1) Qualitatively, we confirmed that all tensors obey the symmetry constraints imposed by their corresponding irreducible representations and (2) Quantitatively, we compared each polarizability tensor component obtained from the moment-based scheme with the corresponding values from the energy-based scheme. The ground electronic and nuclear configuration of molecules can be segregated into point groups depending on its active symmetry elements. There are 32 point groups in total, out of which 19 point groups are of common occurrence in nature; these are the $C_{1}$, $C_{\infty v}$, $C_{2}$, $C_{2v}$, $C_{2h}$, $C_{3v}$, $C_{3h}$, $C_{4v}$, $C_{s}$, $C_{i}$, $D_{2d}$, $D_{2h}$, $D_{3d}$, $D_{3h}$, $D_{6h}$, $D_{\infty h}$, $O_{h}$, $T_{d}$ and $K_{h}$ point groups. An example molecule from each point group is considered and the polarizability tensors ($\alpha$, $A$ and $C$) are calculated using the PBE and BLYP functionals. The tensors are calculated via the energy-based scheme (using G16) and moment-based scheme (using C-Pol implementation in GPAW). The values calculated using G16 are subject to a basis set error due to the finite number of atomic orbitals and are therefore extrapolated to the (energy-based) complete basis set limit (see Appendix \ref{app:CBS}) following the energy-perturbation scheme proposed by Lobota et al \cite{Loboda2016}. This results in 763 unique SCF calculations (using the energy-based scheme) for each system calculated, if the underlying molecular symmetry is not taken into consideration. The values calculated using the momentum-based scheme applied to GPAW require extrapolation to the zero perturbation limit to obtain the converged asymptote (see Methods section: \ref{method:ex_scheme}), and a low enough grid-spacing for convergence with respect to the real-space grid mesh (see e.g. Figure~\ref{fig:timing}). The extrapolation scheme is demonstrated using the H$_2$O (with $C_{2v}$ symmetry) molecule and is presented in Figure \ref{fig:extrapolate}, where the non-zero unique polarizability tensor components converge to a constant value (as indicated in Figure \ref{fig:extrapolate} (i) and the components which should be zero, as mandated by the $C_{2v}$ point group symmetry of H$_2$O, converge to zero (Figure \ref{fig:extrapolate} (ii)). The polarizability tensors can be calculated using the moment-based scheme in just 28 SCF calculations; this includes a single ground state calculation and nine perturbation calculations (the perturbations being made using fields: $V_x$, $V_y$,$V_z$ and field gradients:  $V_{xx}$, $V_{xy}$, $V_{xz}$, $V_{yy}$, $V_{yz}$, $V_{zz}$) for each of the three magnitudes of the perturbation (as mentioned in the extrapolation scheme \ref{method:ex_scheme}).

The moment-based (GPAW) and energy-based (G16) polarizability tensor values for the different point groups are provided in Section 6 of the SI tables S2-S21, and it is shown that the non-zero unique tensor elements conform to the symmetry mandated by the point group of the molecule. An overview comparing GPAW and G16 values is presented in Figure S6 of the SI , where the maximum relative discrepancy (see Section 6 in SI) is used as a measure of agreement between the components.

A comparison between the $\alpha_{x,x}$, $A_{z,zz}$ and $C_{xx,xx}$ polarizability tensor elements is done for a select few examples extracted from the SI tables S1-S20 and is presented in Figure \ref{fig:compPol}. 

\begin{figure}[!ht]
  \centering

  \begin{subfigure}[b]{0.23\textwidth}
    \centering
    \includegraphics[width=\linewidth]{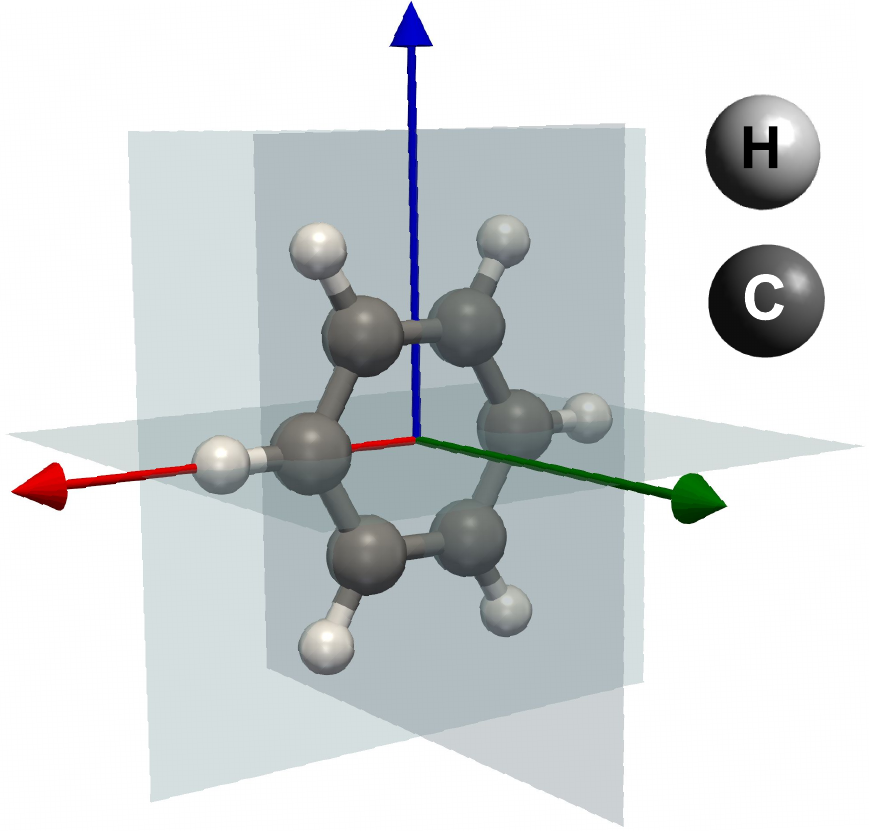}
    \caption{$C_{6}H_{6}$.}
    \label{fig:c6h6-alone}
  \end{subfigure}
  \hfill
  \begin{subfigure}[b]{0.75\textwidth}
    \centering
    \small
    \rowcolors{2}{gray!25}{white}
    \renewcommand{\arraystretch}{1.3}
    \begin{tabular}{|p{0.12\linewidth}|p{0.16\linewidth}|p{0.16\linewidth}|p{0.16\linewidth}|p{0.16\linewidth}|}
      \rowcolor{blue!30}
      \hline
      \textbf{Tensor} & \textbf{G16} & \textbf{GPAW} & \textbf{PySCF} & \textbf{NWChem} \\
      \hline
      $\alpha_{x,x}$ & 84.024  & 85.307  & 84.649 & 82.502 \\
      $\alpha_{y,y}$ & 45.261  & 45.231  & 44.759 & 44.206 \\
      $C_{xx,xx}$    & 663.641 & 667.411 & 680.515 & 660.833 \\
      $C_{xy,xy}$    & 293.805 & 293.518 & 287.440 & 288.270 \\
      $C_{yy,yy}$    & 288.406 & 284.898 & 266.907 & 279.022\\
      \hline
    \end{tabular}
    \caption{Non-zero dipole-dipole ($\alpha$) and quadrupole-quadrupole ($C$) tensor components of benzene.}
    \label{tab:c6h6}
  \end{subfigure}

   \caption{
  (a) The orientation of $C_{6}H_{6}$ used for the tensor calculations, and (b) comparison between energy-based (G16) and moment-based (GPAW, NWChem, and PySCF) calculated polarizability tensors ($\alpha$ ($a_{0}^3$), $A$
  ($a_{0}^4$), and $C$ ($a_{0}^5$) tensors). For $C_{6}H_{6}$ molecule there are 2 ($\alpha$), 0 ($A$), and 3 ($C$) unique elements  respectively, for the $D_{6h}$ point group~\cite{buckingham}.}
  \label{fig:c6h6-combined}
\end{figure}

\subsection{Code comparison}

To demonstrate the correct functioning of the \texttt{C-Pol} implementation in \texttt{NWChem} and \texttt{PySCF} (as described in Section~\ref{method:implementation}), we report in Table~\ref{fig:c6h6-combined} the non-zero unique polarizability tensor components for the benzene molecule, $C_{6}H_{6}$ with
$D_{6h}$ symmetry, in the standard orientation shown in
Figure~\ref{fig:c6h6-alone}. The table \ref{tab:c6h6} collects results from the
moment-based \textsc{C-Pol} implementation (in \textsc{NWChem}, \textsc{PySCF},
and \textsc{GPAW}) and from an energy-based scheme (using finite-field energy calculations
with \textsc{G16}).
Table~\ref{tab:c6h6} confirms both qualitative and quantitative agreement:
for the $D_{6h}$ point group the $\alpha$, $A$, and $C$ tensors possess
exactly 2, 0, and 3 non-zero unique components, respectively, in line with
group-theoretical expectations~\cite{buckingham}.


\noindent
\begin{figure}[!th]
    \begin{center}
    \includegraphics[width=0.48\paperwidth]{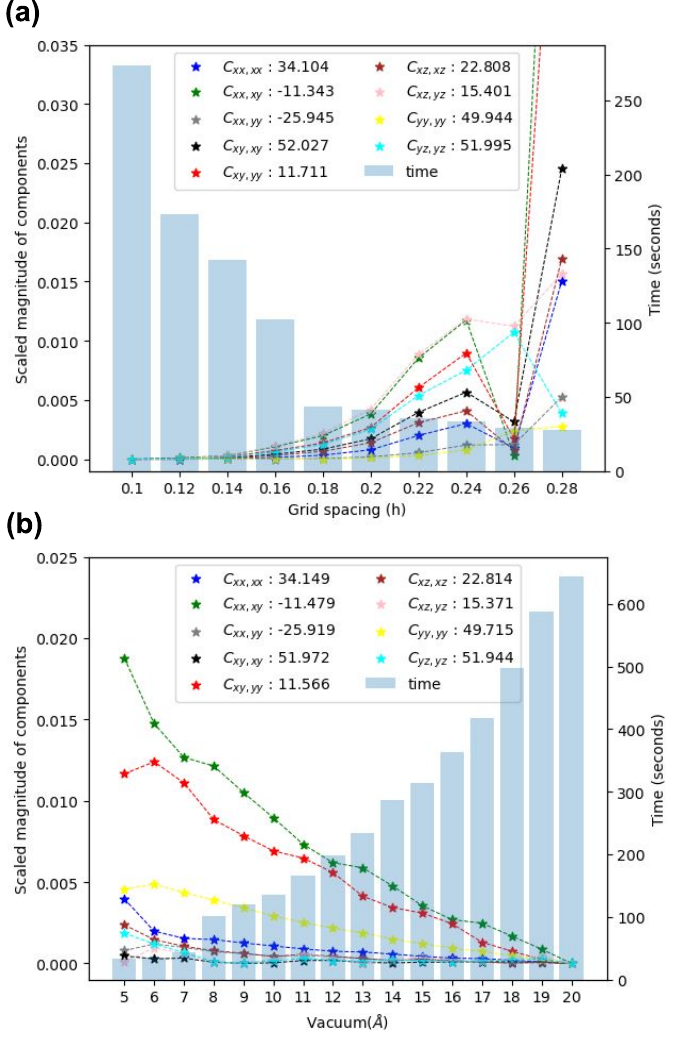}        
    \end{center}
    \caption{\justifying{ Convergence tests for the C-Pol polarizability tensor ($C_{ij,kl}$) components implementation in GPAW as a function of key computational parameters. (a) $C_{ij,kl}$ component values (GPAW) and computational cost as a function of
real-space grid spacing ($h$), indicating convergence for $h \leq 0.18$~\AA,
with rapidly increasing cost at finer grids. (b) Variation of $C_{ij,kl}$ components (GPAW) as a function of vacuum layer
thickness (\AA), with the associated computational time (bar chart, right axis)
increasing steeply beyond $\sim$15~\AA; components stabilize around 10-12~\AA\
of vacuum.}}
    \label{fig:timing}
\end{figure}

\begin{figure}[!th]
    \begin{center}
    \includegraphics[width=0.48\paperwidth]{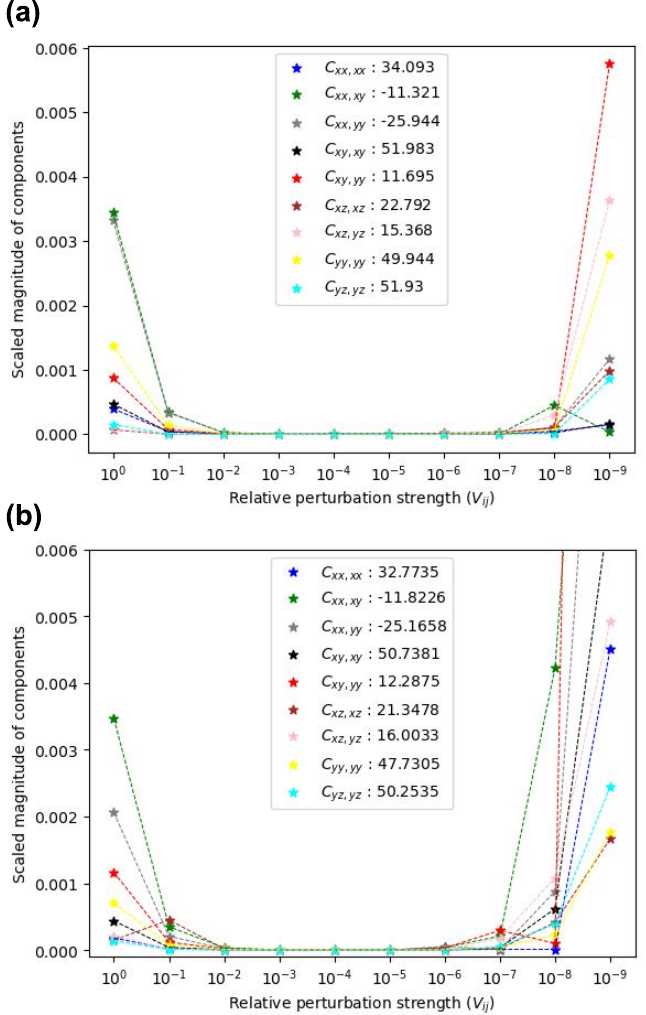}        
    \end{center}
    \caption{\justifying{Convergence of $C_{ij,kl}$ tensor components (calculated using (a) GPAW and (b)PySCF) with respect to relative perturbation strength ($V_{ij}$) (a) indicates a convergence
window around $10^{-2}$ to $10^{-6}$ for GPAW, (b) demonstrates robust convergence in the range
$10^{-4}$ to $10^{-7}$ for the PySCF implementation.}}
    \label{fig:epert}
\end{figure}

\subsection{Convergence and basis set extrapolation}

Convergence studies were performed for the GPAW calculations to study the effect of the resolution of the real-space grid, the perturbation strength (field or field gradient), and the size of the vacuum region on the calculated polarizability tensors, as illustrated in Figures \ref{fig:epert}-\ref{fig:timing} (and in more detail in the Supporting Information Figures. S2, S3, S4 and S5). All convergence tests were performed on a H$_2$O$_2$ molecule. A grid spacing of 0.18~\AA\ was found to provide a suitable compromise between accuracy and computational efficiency. At this spacing the calculated value for the quadrupole-quadrupole polarizability are well converged, as shown in Figure \ref{fig:timing} a. The discrepancy between the calculated value relative to the calculated value at 0.1 \AA\ spacing differs by less than 0.5\% for all but one component, $A_{z,xx}$. The $A_{z,xx}$ tensor component (Figure S3 (b) in SI)  is numerically much smaller than the other components (only about 5\% compared to the next smallest component), hence it is more sensitive to the spatial resolution. However, it is clear that such numerically small components can be well converged by increasing the resolution. The components of the $\alpha$ tensor and other $A$ tensor components are similarly well converged (see the Section 5.1 in the SI). Similar to the grid-spacing convergence tests, we have done convergence tests for the vacuum spacing assigned (Figure \ref{fig:timing}b and section 5.2 in SI) and the relative field perturbation magnitude (Figure \ref{fig:epert}a and section 5.3 in SI). From these we have concluded that a vacuum of 7.0 \AA  \:and a relative perturbation range of $10^{-2}$ to $10^{-7}$ is an optimal choice for accuracy and efficiency.

The other implementations of C-Pol on quantum chemistry codes (NWChem and PySCF) which use basis-sets instead of a finite-difference grid (like GPAW) were also subjected to convergence tests (\ref{fig:epert}b and section 5.4). The H$_2$O$_2$ example was used with the pyscf-cpol implementation for the perturbation strength and basis set convergence tests.

\subsection{Comparison to analytic method}

The polarizability tensor components ($\alpha_{xx}$, $\alpha_{yy}$, $\alpha_{zz}$) were also computed using analytical Coupled-Perturbed Hartree-Fock approach (CPHF) for accuracy and computational efficiency comparisons with the finite-difference (QMMM) approach. The tensor components of the dipole-dipole polarizability ($\alpha$) calculated using finite difference are in excellent agreement with those obtained from the analytic (CPHF) method across all three basis sets, with maximum deviations of less than 0.02 a.u. at each level of theory, confirming that the finite-difference scheme introduces no meaningful numerical error relative to the analytic response (as given in Table \ref{Tab:h2ocphf}). 

\begin{table}[!ht]
    \centering
    \rowcolors{2}{gray!25}{white}
    \renewcommand{\arraystretch}{1.5} 
    \begin{tabular}{|l|c|c|c|c|}
        \hline
        \textbf{Psi4(QMMM)} & \textbf{PVDZ} & \textbf{PVTZ} & \textbf{PVQZ} & \textbf{CBS}\\
        \hline
        $\alpha_{x,x}$ & 8.0031 & 8.4164 & 8.5465 & 8.6414\\
        $\alpha_{y,y}$ & 12.5130 & 12.6664 & 12.6964 & 12.7183\\
        $\alpha_{z,z}$ & 10.0330 & 10.3831 & 10.4864 & 10.5618\\
        Time(secs) & 1.222 & 1.738 & 3.951 & 6.911\\
        \hline
        \textbf{Psi4(CPHF)} & \textbf{PVDZ} & \textbf{PVTZ} & \textbf{PVQZ} & \textbf{CBS}\\
        \hline
        $\alpha_{x,x}$ & 8.0156 & 8.4249 & 8.5559 & 8.6515\\
        $\alpha_{y,y}$ & 12.5036 & 12.6592 & 12.6903 & 12.7130\\
        $\alpha_{z,z}$ & 10.0415 & 10.391 & 10.4916 & 10.5650\\
        Time(secs) & 0.572 & 1.448 & 3.353  & 5.373\\
        \hline
    \end{tabular}
    \caption{Comparison of polarizability tensor components ($\alpha_{x,x}$, $\alpha_{y,y}$, $\alpha_{z,z}$) and computation time for H$_2$O using Psi4 with QMMM and CPHF methods across different basis sets (aug-cc-pVDZ, aug-cc-pVTZ, and aug-cc-pVQZ).}
    \label{Tab:h2ocphf}
\end{table}

While the CPHF method shows a marginal timing advantage at smaller basis sets (0.572 s vs. 1.222 s at aug-cc-pVDZ), the two approaches converge toward comparable wall times as the basis set increases, for example 1.448 s vs. 1.738 s at aug-cc-pVTZ and 3.353 s vs. 3.951 s at aug-cc-pVQZ and reach the same complete basis set (CBS) limit within numerical precision of the extrapolation. This convergence in timings is significant because it highlights a key practical advantage of the finite-difference approach: unlike CPHF, which requires the iterative solution of response equations in the molecular orbital basis with formal O($N^3$) scaling, the finite-difference method relies solely on a small number of standard energy evaluations under applied finite fields, with no additional integral transformation or response solver overhead. As a consequence, CPHF becomes increasingly prohibitive for larger molecules due to the storage and transformation of two-electron integrals in the molecular orbital basis, which introduces severe memory and I/O bottlenecks that are difficult to mitigate even with density-fitting approximations \cite{cholsky}. The finite-difference approach carries no such overhead and scales simply with the cost of a single-point energy calculation, making it the more tractable and transferable choice for larger QM regions, higher-order correlated methods, and QM/MM frameworks where analytic CPHF responses are unavailable or impractical to implement.

\subsection{Electrostatic potential plots}

The electrostatic potential is projected on to an iso-surface corresponding to the volume encompassing 95\% of the electronic density and is calculated using both GPAW ($V_{QM}$), and from the Buckingham expansion employing the moments and polarizabilities derived from GPAW ($V_{MM}$). The self-consistent field equations for the induced moments and resulting electrostatic potential are presented in Appendix~\ref{app:SCF}. $V_{QM}$ and $V_{MM}$ along with their difference ($\Delta V$) for H$_2$O and CO$_2$ dimer systems are presented in Figures 6 (i) and (ii).

\begin{figure}[!th]
    \begin{center}
    \includegraphics[width=12.0cm]{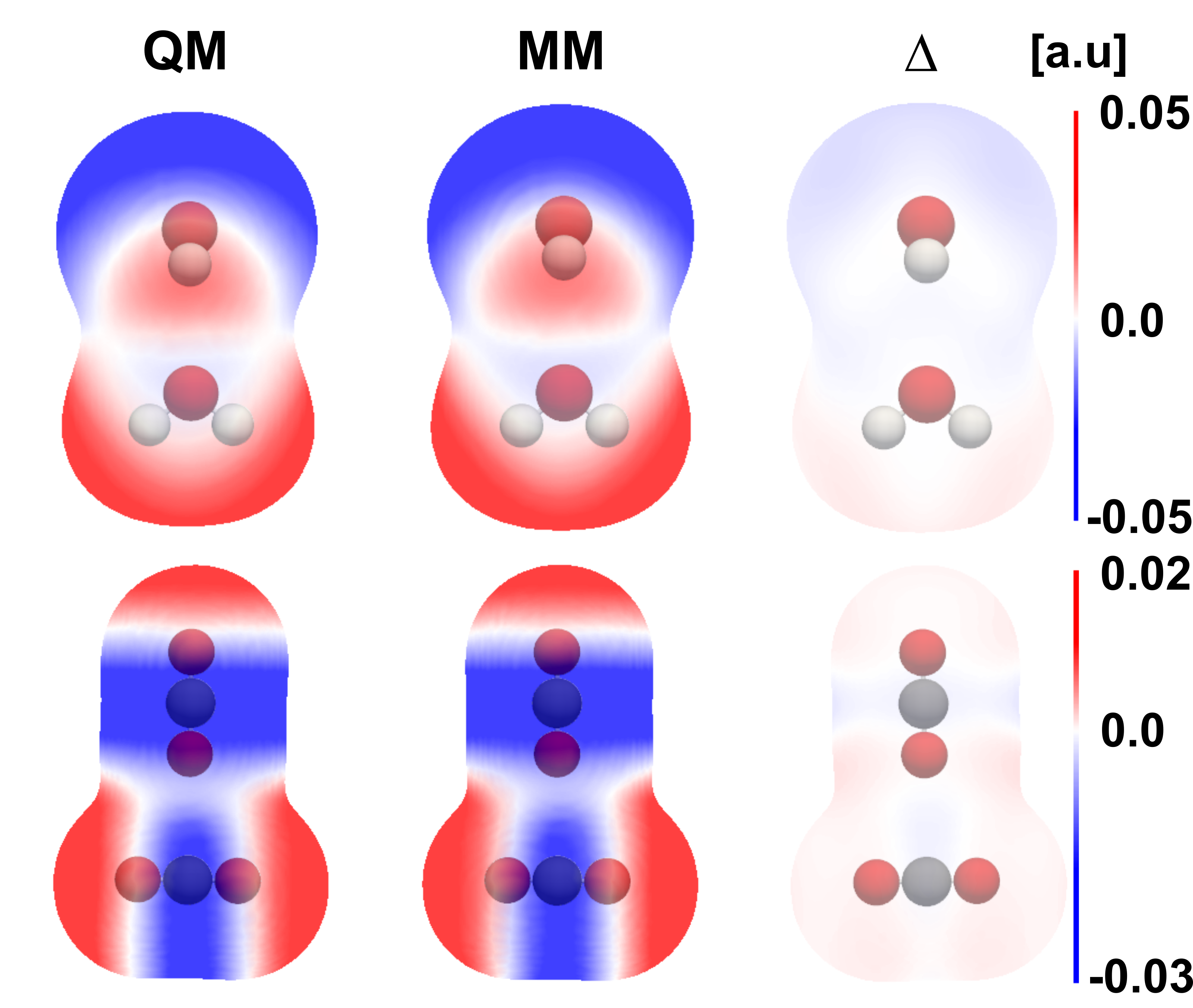}  
    \end{center}
    \caption{\justifying{The calculated moments and polarizabilities using the moment-based method are used to evaluate the semi-classical potential, $V_\mathrm{MM}$, and in comparison to the QM potential, $V_\mathrm{QM}$, on a fixed iso-surface representing 95\% of the charge density for a dimer system consisting of two H$_2$O molecules and two CO$_2$ molecules, top and bottom respectively. The potential difference between the classical and quantum description is shown on the right, $\Delta V$.}}
    \label{fig:potsurf}
\end{figure}

The low discrepancy between the QM and MM potential energy surfaces indicates that the underlying QM interactions are accurately captured by the calculated distributed moments and polarizabilities. This agreement is particularly pronounced for the CO$_2$ dimer, where the deviation between QM and MM potentials is effectively negligible at the considered volumetric distance. Quantitatively, the differences remain very small, with a maximum deviation of 0.0013, a minimum of -0.0019, a mean absolute error (MAE) of 0.0003, and a root-mean-square (RMS) deviation of 0.0006. These results highlight the robustness of the electrostatic and polarization representation for this system. For the H$_2$O dimer, slightly larger but still modest deviations are observed, reflecting the more complex and anisotropic hydrogen-bonding interactions present in water. The maximum and minimum deviations are 0.0031 and -0.0052, respectively, with an MAE of -0.0009 and an RMS deviation of 0.0026. Despite these increased deviations relative to CO$_2$, the overall agreement remains good (under 10\% max discrepancy). Overall, these results demonstrate that the chosen representation of molecular moments and polarizabilities provides a consistent and accurate description of intermolecular interactions across both weakly interacting and hydrogen-bonded dimers, reinforcing its suitability for modeling condensed-phase and cluster environments.

\section{Summary and Conclusion}

We have presented C-Pol, a lightweight and backend-agnostic Python package for computing multipole polarizability tensors through a point-charge perturbation scheme. By replacing conventional applied-field perturbations with simple QM/MM-style point charge arrangements, the method produces the electric field and field-gradient perturbations required to evaluate the dipole-dipole ($\alpha$), dipole-quadrupole (A), and quadrupole-quadrupole (C) polarizability tensors from finite differences of multipole moments, rather than from energy differences. This design choice yields a roughly 30-fold reduction in the number of required SCF calculations relative to energy-based complete-basis-set schemes, while delivering numerical agreement within ~3\% across molecules spanning 19 common point groups. Because C-Pol interfaces with any quantum chemistry code that supports external point charges (a feature ubiquitous in QM/MM implementations), it imposes no restrictions on the choice of electronic structure method, basis set, or code, making it straightforwardly extensible to new back-ends.

Several directions offer natural extensions of this work. The point-charge perturbation formalism can in principle be generalized to field gradients of arbitrary rank, enabling computation of higher-order polarizabilities such as the dipole-octupole (D) tensor; the requisite charge distributions are outlined in the Supporting Information. Integration with a broader range of open-source codes (including periodic DFT packages) would further widen applicability, particularly for condensed-phase and surface systems. The efficiency of the method makes it well-suited to generating large, high-quality training datasets for machine-learning interatomic potentials and response-property models, where systematic coverage of chemical and conformational space demands thousands of polarizability evaluations. Finally, the zero-perturbation extrapolation scheme naturally extends to mapping polarizability tensors as functions of molecular geometry, providing a rigorous route to flexible, geometry-dependent polarizable force fields of the kind required by models such as SCME \cite{EOJ2022} and polarizable embedding QM/MM frameworks \cite{eoj2019polar1,eoj2019polar2,bessner2026}.

\section*{Acknowledgements}


\paragraph{Funding information}
This work was supported by the Icelandic Research Fund (grant no. 2410644-051) and the Eimskip fund (grant no. HEI2023-93153). Computer resources, data storage and user support were provided by the Division of Information Technology of the University of Iceland through the Icelandic Research e-Infrastructure (IREI) project, funded by the Icelandic Infrastructure Fund. The authors acknowledge the support from the state of Baden-W\"urttemberg through bwHPC and the German Research Foundation (DFG) through grant no INST 40/575-1 FUGG (JUSTUS 2 cluster), as well as support from the BMBF (Bundesministerium fur Bildung und Forschung) through the project CASINO (FKZ: 03XP0487G).

\section*{Supporting Information}
 The supporting information has five sections:  (i) The tracelessness conditions that should be satisfied by the polarizabilty tensors. (ii) Derivation of field and field-gradient tensors for different charge distributions. (iii) Properties of charge distributions which can be used to generalize the method to higher order gradients. (iv) Convergence tests of the polarizabilities with respect to the parameters of the reals-space grid based wavefunctions. (v) Polarizabilities (and symmetries) of all of the molecules considerd in this work. The python module and wrapper for GPAW, PySCF and NWChem is freely available \href{https://anoopanair.gitlab.io/hmpol-docs/index.html}{online}. The potential isosurfaces shown in Figure~\ref{fig:potsurf} are drawn using ChimeraX \cite{chimeraX}, and the molecules and axis are drawn in PyVista \cite{pyvista}.

\begin{appendix}
\numberwithin{equation}{section}
\section{Traceless moments in GPAW}
\label{app:GPAW}

Details on the projector augmented wave method (PAW) \cite{Blochl1994} and how pseudo electron states are represented in GPAW can be found elsewhere \cite{gpaw,gpaw2,gpaw3}.
We make use of the pseudo charge density of GPAW which is given by
\begin{equation}
 \rho(\mathbf{r}) = \tilde{n}(\mathbf{r}) + \sum_a\tilde{Z}^a(\mathbf{r})
\end{equation}
where $\tilde{n}(\mathbf{r})$ is the pseudo valence electron density and $\tilde{Z}^a$ are atom-centred compensation charges. The atom-centred compensation charges are expanded in terms of real-space solid spherical harmonics
\begin{equation}
    \tilde{Z}^a(\mathbf{r}) = \sum_L \Delta^a_{L\alpha\beta}D^a_{\alpha\beta}\tilde{g}_L^a(\mathbf{r})
\end{equation}
where $D^a_{\alpha\beta}$ are one-centered density matrices (expansion coefficients of the difference between $\alpha\beta$ atom-centered pseudo and all-electron partial waves), and $\Delta^a_{L\alpha\beta}$ are multipole expansion coefficients of rank $L$. The pseudo functions are represented on a domain decomposed regular real space grid making the evaluation of the integrals trivial. The traceless moments in terms of the pseudo charge density are
\begin{align}
 \mu_\alpha =& \int \rho(\mathbf{r})r'_\alpha d\mathbf{r} \\
 \theta_{\alpha\beta} =& \frac{1}{2}\int \rho(\mathbf{r})\left(3r'_\alpha r'_\beta -|\mathbf{r}'|^2\delta_{\alpha\beta}\right) d\mathbf{r} \\
 \Omega_{\alpha\beta\gamma} =& \frac{1}{6}\int \rho(\mathbf{r})\big(
 15r'_\alpha r'_\beta r'_\gamma \nonumber \\
 &-3|\mathbf{r}'|^2\left(r'_\alpha\delta_{\beta\gamma} + r'_\beta\delta_{\alpha\gamma} + r'_\gamma\delta_{\alpha\beta}\right)\big) d\mathbf{r} \\
 \Phi_{\alpha\beta\gamma\delta} =& \frac{1}{24}\int \rho(\mathbf{r})\big(105 r'_\alpha r'_\beta r'_\gamma r'_\delta \nonumber \\
 &-15|\mathbf{r}'|^2\big(r'_\alpha r'_\beta \delta_{\gamma\delta} + r'_\alpha r'_\gamma \delta_{\beta\delta} + r'_\alpha r'_\delta \delta_{\beta\gamma} \nonumber \\
 & \ \ \ \ \ \ \ \ \ \ \ \ \ 
 r'_\beta r'_\gamma \delta_{\alpha\delta} +
 r'_\beta r'_\delta \delta_{\alpha\gamma} + 
 r'_\gamma r'_\delta \delta_{\alpha\beta} \big) \nonumber \\ 
 &+ 3|\mathbf{r}'|^4\big(\delta_{\alpha\beta}\delta_{\gamma\delta} + 
 \delta_{\alpha\gamma}\delta_{\beta\delta} + 
 \delta_{\alpha\delta}\delta_{\beta\gamma}\big)\big)d\mathbf{r} 
\end{align}
where $\mathbf{r}' = \mathbf{r} - \mathbf{R}$, and $\mathbf{R}$ is a some choice of origin, such as the center of mass of the molecule of interest.

\section{Complete basis set limit}
\label{app:CBS}

The Complete Basis Set (CBS) Method is used to estimate the energy of a molecular system at the limit of an infinitely large basis set. 

\textbf{Exponential CBS scheme for energies: }In quantum chemistry calculations, the wavefunction of a system is approximated using basis sets of finite size, and the energy values converge to an asymptote as the basis set size increases. The exponential extrapolation scheme as given in equation \ref{eq:exp_cbs} is used to perform CBS\cite{CBS}. 

\begin{align}
    E(X) = E_{\infty} + Be^{-\alpha X} \label{eq:exp_cbs}
\end{align}

Where $E_{\infty}$ is the converged energy asymptote , $E(X)$ is the energy due to basis set X and $\alpha$, $B$ are the fitted parameters. For the basis sets aug-cc-pVDZ (DZ), aug-cc-pVTZ (TZ) and aug-cc-pVQZ (QZ) equation \ref{eq:exp_cbs} becomes the set of equations given in.\ref{eq:sys_cbs}.

\begin{align}
    E(DZ) &= E_{\infty} + Be^{-2\alpha} \\
    E(TZ) &= E_{\infty} + Be^{-3\alpha} \\
    E(QZ) &= E_{\infty} + Be^{-4\alpha} \label{eq:sys_cbs}
\end{align}

The system of equations in \ref{eq:sys_cbs} can be solved to obtain $E_{\infty}$ as in equation \ref{eq:einf}.

\begin{align}
    E_{\infty}  = \frac{E(DZ)E(QZ) - E(TZ)^2}{E(DZ)-2E(TZ)+E(QZ)}\label{eq:einf}
\end{align}

\textbf{Helgaker CBS equation for multipoles:} For moments and polarizability tensors that exhibit slow basis-set convergence, we adopt the Helgaker two-point extrapolation scheme to obtain CBS estimates.
\begin{equation}
  P_{\mathrm{CBS}} =
  \frac{P_X (X-1)^3 - P_{X-1} X^3}{(X-1)^3 - X^3},
  \label{eq:Helgaker-property}
\end{equation}
where $P_X$ and $P_{X-1}$ are the values of the property computed with
basis sets of cardinal numbers $X$ and $X-1$, respectively, and
$X = 2, 3, 4, \ldots$ for the D, T, Q, \ldots\ members of the series.
This expression assumes an asymptotic $X^{-3}$ convergence of the
basis-set error and has been shown to provide reliable CBS limits for a
wide range of molecular properties \cite{helgaker}.

\section{Self-consistent field equations}
\label{app:SCF}
The self-consistent field equations of SCME~\cite{Wikfield2013,EOJ2022,Myneni2022} are as follows:
given the potential field, $V^{i}_{\alpha}$, and the potential field gradient,
$V^{i}_{\alpha\beta}$, at the COM of molecule $i$, the molecules are polarized
resulting in induced dipole and quadrupole moment
\begin{align}
\Delta \mu^{i}_{\alpha} &= - \alpha^{i}_{\alpha,\beta} V^{i}_{\beta}
- \frac{1}{3} A^{i}_{\alpha,\beta\gamma} V^{i}_{\beta\gamma},
\label{eq:C23} \tag{23} \\
\Delta \theta^{i}_{\alpha\beta} &= - A^{i}_{\gamma,\alpha\beta} V^{i}_{\gamma}
- C^{i}_{\gamma\delta,\alpha\beta} V^{i}_{\gamma\delta}
\label{eq:C24} \tag{24}
\end{align}
where the external field is given by
\begin{equation}
V^{i}_{\alpha} = \sum_{j \ne i}^{n} V^{ij}_{\alpha},
\label{eq:C25} \tag{25}
\end{equation}
and the contribution to the external field at site $i$ due to site $j$ is given by
\begin{align}
V^{ij}_{\alpha} &=
- T^{ij}_{\alpha\beta} \bigl(\mu^{j}_{\beta} + \Delta \mu^{j}_{\beta}\bigr)
+ \frac{1}{3} T^{ij}_{\alpha\beta\gamma} \bigl(\theta^{j}_{\beta\gamma}
+ \Delta \theta^{j}_{\beta\gamma}\bigr) \notag \\
&\quad - \frac{1}{15} T^{ij}_{\alpha\beta\gamma\delta} \Omega^{j}_{\beta\gamma\delta}
+ \frac{1}{105} T^{ij}_{\alpha\beta\gamma\delta\eta} \Phi^{j}_{\beta\gamma\delta\eta}
\label{eq:C26} \tag{26}
\end{align}
The field gradient and higher order gradients are given by the subsequent
use of the gradient operator, for example, $\nabla_{\beta} V^{i}_{\alpha}
= V^{i}_{\alpha\beta}$, $\nabla_{\gamma} V^{i}_{\alpha\beta}
= V^{i}_{\alpha\beta\gamma}$.
$T^{ij}$ is the zeroth-order Coulomb interaction tensor
$T^{ij} = 1/r^{ij}$ where $r^{ij}$ is the distance between molecule $i$ and $j$.

With a SCF solution of the induced moments the potential at any point can be
generated using
\begin{align}
V_{\mathrm{MM}}(\mathbf{r}) &= \sum_{i}
\Bigl[
- T^{ri}_{\alpha} \bigl(\mu^{i}_{\alpha} + \Delta \mu^{i}_{\alpha}\bigr)
+ T^{ri}_{\alpha\beta} \bigl(\theta^{i}_{\alpha\beta}
+ \Delta \theta^{i}_{\alpha\beta}\bigr) \notag \\
&\qquad\quad
- T^{ri}_{\alpha\beta\gamma} \Omega^{i}_{\alpha\beta\gamma}
+ T^{ri}_{\alpha\beta\gamma\eta} \Phi^{i}_{\alpha\beta\gamma\eta}
\Bigr],
\label{eq:C27} \tag{27}
\end{align}
where $ir$ refers to the distance between the COM of molecule $i$ and grid
point $\mathbf{r}$ in a regular grid.

For comparison we extract the electrostatic potential from GPAW using built-in
routines and calculate the electrostatic potential difference as
\begin{equation}
\Delta V(\mathbf{r}) = V_{\mathrm{QM}}(\mathbf{r}) - V_{\mathrm{MM}}(\mathbf{r}).
\label{eq:DeltaV}
\end{equation}

\end{appendix}

\bibliographystyle{SciPostBib}
\bibliography{main.bib}

\end{document}


\tableofcontents
\clearpage

\section{Traceless tensors}

The  Traceless-ness condition for the dipole-quadrupole polarizability is represented as: 
\begin{align}
    \Sigma_{j=\{x,y,z\}} \:A^{'}_{i,jj} &= 0, \forall \: i \: \epsilon \: \{x,y,z\}
\end{align}

To convert the traced dipole-quadrupole polarizability ($A^{o}$) to the traceless one ($A^{'}$), the equations given below can be used. 

\begin{align}
A^{'}_{i,ii} &= 2A^{o}_{i,ii} - [A^{o}_{i,jj} + A^{o}_{i,kk}] , \mathrm{where }\: i\neq j \neq k\\
A^{'}_{i,jj} &= 2A^{o}_{i,jj} - [A^{o}_{i,ii} + A^{o}_{i,kk}],  \mathrm{where }\: i\neq j \neq k\\
A^{'}_{i,jk} &= \frac{3}{2}A^{o}_{i,jk}, \mathrm{where }\: j \neq k
\end{align}

The Traceless-ness condition for the quadrupole-quadrupole polarizability is represented as: 
\begin{align}
    \Sigma_{k=\{x,y,z\}}\:C^{'}_{ij,kk} &= 0, \forall \: i, \:j \: \epsilon \: \{x,y,z\}
\end{align}

To convert the traced quadrupole-quadrupole polarizability ($C^{o}$) to the traceless one ($C^{'}$), the equations given below can be used. 

\begin{align}
C^{'}_{ii,ii} &= \frac{1}{3}\{ 4C^{o}_{ii,ii} + C^{o}_{jj,jj} + C^{o}_{kk,kk}
                \\\nonumber
              & - 4C^{o}_{ii,jj} - 4C^{o}_{ii,kk} + 2C^{o}_{jj,kk}
              \} ,  \mathrm{where }\: i\neq j \neq k
\end{align}
\begin{align}
C^{'}_{ii,jj} &= \frac{1}{3}\{ -2C^{o}_{ii,ii} -2C^{o}_{jj,jj} + C^{o}_{kk,kk}
                \\\nonumber
              & + 5C^{o}_{ii,jj} - C^{o}_{ii,kk} - C^{o}_{jj,kk}
              \} ,  \mathrm{where }\: i\neq j \neq k
\end{align}
\begin{align}
C^{'}_{ii,jk} &= \frac{1}{2}\{ 2C^{o}_{ii,jk} - C^{o}_{kk,jk} - C^{o}_{jj,jk}\},  \mathrm{where }\: j \neq k
\end{align}
\begin{align}
C^{'}_{ij,ik} &= \frac{3}{4}\{ C^{o}_{ij,ik} \},  \mathrm{where }\: i\neq k
\end{align}

\section{Deriving the field and field-gradient equations}

\begin{figure}[!ht]
    \centering
        \includegraphics[width=0.75\paperwidth]{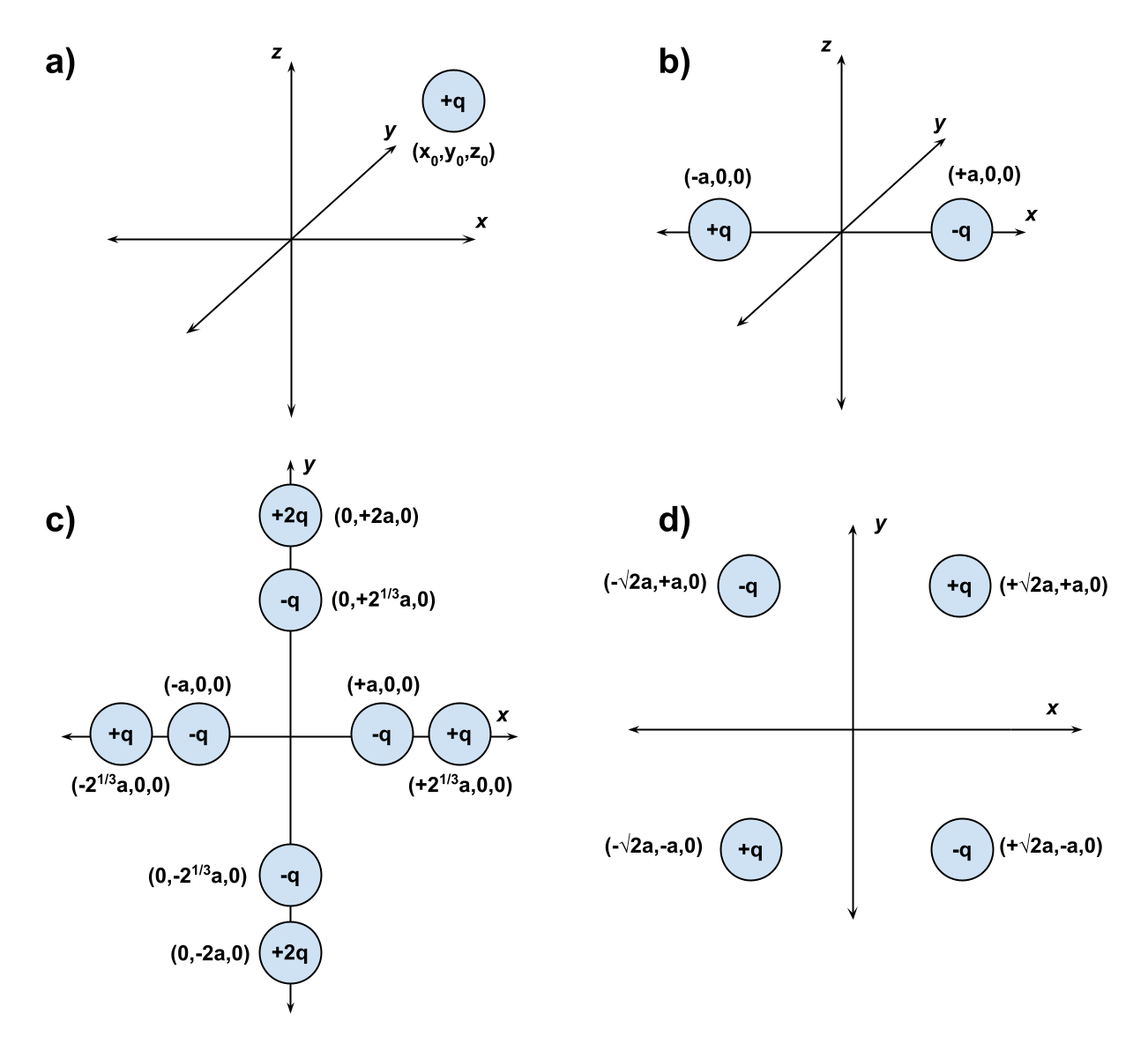}
        \captionof{figure}{(a) Example point charge. (b) Point charge distribution (PCD) to produce a field of $V_{x}$ = ${2q}.{a}^{-2}$. (c) The PCD produces a field-gradient of $V_{xx}$ =  $3q.(2a)^{-3}$. (d) The PCD produces a field-gradient of $V_{xy}$ = $4\sqrt{2}q(\sqrt{3}a)^{-3}$.}
    \label{fig:pc}
\end{figure}

The coulombic potential, its first and second order derivative indicating the fields and field gradients are represented in equations  \ref{Eq:pot}, \ref{Eq:potF} and \ref{Eq:potFF} respectively. Here $r = \sqrt{x_{0}^2+y_{0}^2+z_{0}^2}$.

\begin{align}
   V(q,x_{0},y_{0},z_{0}) &= \frac{q}{\sqrt{x_{0}^2+y_{0}^2+z_{0}^2}}\label{Eq:pot}
\end{align}

\begin{align}
   V^{'}(q,x_{o}, y_{o}, z_{o}) &= \begin{bmatrix}
         \frac{\partial V}{\partial x}\\
        \frac{\partial V}{\partial  y}\\
        \frac{\partial V}{\partial  z}
        \end{bmatrix}
     = \begin{bmatrix}
         \frac{-qx_{o}}{r^3}\\
        \frac{-qy_{o}}{r^3}\\
        \frac{-qz_{o}}{r^3}
        \end{bmatrix}\label{Eq:potF}
\end{align}

\begin{align}
   V^{''}(q, x_{o}, y_{o}, z_{o}) &= \begin{bmatrix}
        \frac{\partial V_x}{\partial x}& \frac{\partial V_x}{\partial y}& \frac{\partial V_x}{\partial z}\\
        \frac{\partial V_y}{\partial x}& \frac{\partial V_y}{\partial y}& \frac{\partial V_y}{\partial z}\\
        \frac{\partial V_z}{\partial x}& \frac{\partial V_z}{\partial y}& \frac{\partial V_z}{\partial z}
        \end{bmatrix} \nonumber \\
     &= \begin{bmatrix}
         \frac{-q(-2x_{o}^2+y_{o}^2+z_{o}^2)}{r^5}&\frac{3qx_{o}y_{o}}{r^5}&\frac{3qx_{o}z_{o}}{r^5}\\
         \frac{3qx_{o}y_{o}}{r^5}&\frac{-q(x_{o}^2 -2y_{o}^2+z_{o}^2)}{r^5}&\frac{3qy_{o}z_{o}}{r^5}\\
         \frac{3qx_{o}z_{o}}{r^5}&\frac{3qy_{o}z_{o}}{r^5}&\frac{-q(x_{o}^2+y_{o}^2 -2z_{o}^2)}{r^5}
         \end{bmatrix}\label{Eq:potFF}
\end{align}

The V value of all the charge distributions in figure \ref{fig:pc}
The matrices for each of the charge distributions are as follows: 

\textbf{ Case 1: Creation of a field without a field-gradient }

The charge distribution given in figure \ref{fig:pc}(b) is used for creating a non-zero field-gradient component $-V_{x}$ (where $V_{x}$ is the potential gradient). The potential, and field gradient at the center of the charge distribution will be 0. This is demostrated below: 

\begin{align}
   V(q,a) &= \frac{q}{|a|} + \frac{-q}{|a|} = 0.
\end{align}

\begin{align}
   V^{'}(q,a) &= \begin{bmatrix}
         \frac{-qa}{a^3}  + \frac{-qa}{a^3}      \\
        0         \\
        0
        \end{bmatrix}
        = \begin{bmatrix}
         \frac{-2q}{a^2}      \\
        0         \\
        0
        \end{bmatrix}
\end{align}

\begin{align}
   V^{''}(q, a)  &= \begin{bmatrix}
         \frac{-q(-2a^{2})}{a^5} + \frac{q(-2a^{2})}{a^5} & 0 & 0 \\
         0 & \frac{-q(a^{2})}{a^5} + \frac{q(a^{2})}{a^5} & 0 \\
         0 & 0 & \frac{-q(a^{2})}{a^5} + \frac{q(a^{2})}{a^5}
         \end{bmatrix}\\
         &= \begin{bmatrix}
         0 & 0 & 0 \\
         0 & 0 & 0 \\
         0 & 0 & 0
         \end{bmatrix}
\end{align}
\textbf{Note:} The $-V^{'}(q,a)$ and $-V^{''}(q,a)$ gives the field and field-gradient respectively.

\textbf{ Case 2: Creation of a field-gradient with a zero-field scenario}

The charge distribution given in figure \ref{fig:pc}(c) and (d) are used for creating the $-V_{xx}$ and $-V_{xy}$ field-gradient component. The potential, and field at the center of the charge distribution will be 0. This is demonstrated below following case 1. $V(q,a)$ and $V^{'}(q,a)$ are 0 for both the charge distributions.
For charge distribution in \ref{fig:pc} (c):

\begin{align}
   V^{''}(q, a)  &= \begin{bmatrix}
         \frac{-3q}{2a^3}  & 0 & 0 \\
         0 & 0  & 0 \\
         0 & 0 & \frac{3q}{2a^3}
         \end{bmatrix}\\
\end{align}

Here $V_{zz}$ = $-V_{xx}$ due to the symmetry of the charge distribution. 

For charge distribution in \ref{fig:pc} (d): 
\begin{align}
   V^{''}(q, a)  &= \begin{bmatrix}
         0  & \frac{4\sqrt{2}q}{(\sqrt{3}a)^{3}} & 0 \\
         \frac{4\sqrt{2}q}{(\sqrt{3}a)^{3}} & 0  & 0 \\
         0 & 0 & 0
         \end{bmatrix}\\
\end{align}

Here $V_{xy}$ = $V_{yx}$ due to the symmetry of the charge distribution. 

\textbf{Note:} The $-V^{'}(q,a)$ and $-V^{''}(q,a)$ gives the field and field-gradient respectively.

\section{Properties of the charge distributions}


The point charge distributions conform to a certain symmetry of the charge density moments and as a result have the desired non-zero gradients. This is presented in the table below.

For a point charge distribution to generate a certain (traceless) non-zero perturbation the non-zero moments of the distribution must match the rank of the gradient. For example, a distribution with a non-zero first moment (dipole) generates a non-zero field at the center of the distribution. A distribution with a non-zero second moment (quadrupole) generates at the center of the distribution a non-zero field gradient etc.

The table on the next page lists charge distributions which have a non-zero dipole, quadrupole and octupole, and the corresponding field, field-gradient and higher order gradients associated with each distribution at the center. In order to find the coordinates (i.e. multiplicative factors of the distance metric $a$) for a distribution which has a certain non-zero moment (and zero lower order moments), one can solve a set of equations requiring that lower order fields (and moments) are zero.

In all cases the non-zero gradients of rank $n$ (and ultimately non-zero higher order gradients of rank $n+2$) conform to Laplaces equation $$\sum_{\alpha} \frac{\partial^2 V^n}{\partial \alpha^2} = 0$$

\vfill\eject

\begin{landscape}
\begin{table}[!th]
\begin{tabular}{l|l|l|l|l|l|l|l|l}
PC         & q's      & positions ($\alpha$, $\beta$, $\gamma$)   & V   & V'         & V''                & V'''                                 & V''''                                         & V'''''                                                        \\ \hline
$\mu$      & +q, -q   & (a,0,0), (-a,0,0)                         &     & $V_\alpha$ &                    & $\sum_{\beta}V_{\alpha\beta\beta}=0$ &                                               & $\sum_{\beta}\sum_{\gamma}V_{\alpha\beta\beta\gamma\gamma}=0$ \\ \hline
$\theta_1$ & +q, -q,  & ($\sqrt{2}$a,a,0), ($\sqrt{2}a$,-a,0),    &     &            & $V_{\alpha\beta}$  &                                      & $\sum_{\gamma}V_{\alpha\beta\gamma\gamma}=0$  &                                                               \\
           & +q, -q   & (-$\sqrt{a},-a,0$), (-$\sqrt{2}$a,a,0)    &     &            &                    &                                      &                                               &                                                               \\ \hline
$\theta_2$ & +2q, -q, & (2a,0,0), (2$^{1/3}$a,0,0),               & $V$ &            & $V_{\alpha\alpha}$ &                                      & $\sum_{\gamma}V_{\alpha\alpha\gamma\gamma}=0$ &                                                               \\
           & +q, -q,  & (0,2$^{1/3}$a,0), (0,a,0),                &     &            & $-V_{\beta\beta}$  &                                      & $\sum_{\gamma}V_{\beta\beta\gamma\gamma}=0$   &                                                               \\
           & +2q, -q, & (-2a,0,0), (-2$^{1/3}$a,0,0),             &     &            &                    &                                      &                                               &                                                               \\
           & +q, -q   & (0,-2$^{1/3}$a,0), (0,-a,0)               &     &            &                    &                                      &                                               &                                                               \\ \hline
$\Omega_1$ & +q, -q,  & (a,$\sqrt{3}$,0), (2a,0,0),               &     &            &                    & $V_{\alpha\alpha\alpha}$             &                                               & $\sum_{\beta}V_{\beta\beta\alpha\alpha\alpha}=0$              \\
           & +q, -q,  & (a,-$\sqrt{3}$a,0), (-a,$\sqrt{3}$a,0),   &     &            &                    & $-V_{\gamma\gamma\alpha}$            &                                               & $\sum_{\beta}V_{\beta\beta\gamma\gamma\alpha}=0$              \\
           & +q, -q   & (-2a,0,0), (-a,-$\sqrt{3}$a,0)            &     &            &                    &                                      &                                               &                                                               \\ \hline
$\Omega_2$ & +q, -q,  & (a,0,$\sqrt{3}$a), (0,a,$\sqrt{3}$a),     &     &            &                    & $V_{\alpha\beta\beta}$               &                                               &                                                               \\
           & +q, -q,  & (-a,0,$\sqrt{3}$a), (0,-a,$\sqrt{3}$a),   &     &            &                    & $-V_{\alpha\gamma\gamma}$            &                                               &                                                               \\
           & +q, -q,  & (a,0,-$\sqrt{3}$a), (0,a,-$\sqrt{3}$a),   &     &            &                    &                                      &                                               &                                                               \\
           & +q, -q   & (-a,0,-$\sqrt{3}$a), (0,-a,-$\sqrt{3}$a), &     &            &                    &                                      &                                               &                                                               \\ \hline
$\Omega_3$ & +q, -q,  & (a,a,$\sqrt{3}$a), (-a,a,$\sqrt{3}$a),    &     &            &                    & $V_{\alpha\beta\gamma}$              &                                               & $\sum_{\eta}V_{\alpha\beta\gamma\eta\eta}=0$                    \\
           & +q, -q,  & (-a,-a,$\sqrt{3}$a), (a,-a,$\sqrt{3}$a),  &     &            &                    &                                      &                                               &                                                               \\
           & +q, -q,  & (a,a,-$\sqrt{3}$a), (-a,a,-$\sqrt{3}$a),  &     &            &                    &                                      &                                               &                                                               \\
           & +q, -q,  & (-a,-a,-$\sqrt{3}$a), (a,-a,-$\sqrt{3}$a) &     &            &                    &                                      &                                               &                                                              
\end{tabular}
\label{tbl:properties}
\caption{Description of point charge distributions which have a non-zero dipole ($\mu$), quadrupole ($\theta$) and octupole ($\Omega$), and the corresponding non-zero field, field-gradient and third order gradient at the center of the distribution. The charges and positions of charges are listed in terms of $q$ and spatial metric $a$, respectively.}
\end{table}
\end{landscape}

\section{Energy Based Scheme}

The energy-based scheme represents the polarizability tensors using finite-difference expressions that depend on the system’s energies in the presence of applied fields and field gradients. The different polarizability tensors for a given field ($F_i$/$F_j$) and field-gradient ($F_{ii}$ (diagonal) and $F_{ij}$ (off-diagonal)) are defined as follows:

\textbf{Dipole-dipole polarizability: }

\begin{equation}
\alpha_{ii} F_i^2 =
\frac{5}{2} E(0)
- \frac{4}{3}\bigl[E(F_i) + E(-F_i)\bigr]
+ \frac{1}{12}\bigl[E(2F_i) + E(-2F_i)\bigr]
\label{eq:alpha_ii}
\end{equation}

\begin{equation}
\begin{aligned}
\alpha_{ij} F_i F_j &= \frac{1}{48}\Bigl[
E(2F_i, 2F_j) - E(2F_i, -2F_j)
- E(-2F_i, 2F_j) + E(-2F_i, -2F_j)
\Bigr] \\
&\quad + \frac{1}{3}\Bigl[
E(F_i, F_j) - E(F_i, -F_j)
- E(-F_i, F_j) + E(-F_i, -F_j)
\Bigr]
\end{aligned}
\label{eq:alpha_ij}
\end{equation}

The tensor requires 39 unique energy computations.

\textbf{Dipole–quadrupole polarizability:}

\begin{equation}
P_{ijk} F_i F_{jk}
  = \left(\frac{1}{4}\right)
    \Bigl[
      E(-F_i, F_{jk}) - E(F_i, F_{jk})
      - E(-F_i, -F_{jk}) + E(F_i, -F_{jk})
    \Bigr]
\label{eq:dq_15}
\end{equation}

To obtain the trace-less components the following equations are used.

\begin{equation}
\begin{aligned}
A_{i,ii} &= 2 P_{i,ii} - \bigl[P_{i,jj} + P_{i,kk}\bigr],
\quad \text{with } i,j,k \text{ all different}, \\
A_{i,jj} &= 2 P_{i,jj} - \bigl[P_{i,ii} + P_{i,kk}\bigr],
\quad \text{with } i,j,k \text{ all different}, \\
A_{i,jk} &= \frac{3}{2} P_{i,jk},
\quad \text{with } j \ne k
\end{aligned}
\label{eq:A_traceless}
\end{equation}

There are 72 unique energy computations required for this tensor.

\textbf{Quadrupole–quadrupole polarizability:}

\begin{equation}
P_{ij,ij} F_{ij}^2
  = 2 E(0) - \bigl[E(F_{ij}) + E(-F_{ij})\bigr]
\label{eq:P_ijij}
\end{equation}

\begin{equation}
\begin{aligned}
P_{ij,kl} (F_{ij} F_{kl}) &= \left(\frac{1}{4}\right)
\Bigl[
  E(F_{ij}, F_{kl}) - E(F_{ij}, -F_{kl}) \\
&\quad + E(-F_{ij}, F_{kl}) - E(-F_{ij}, -F_{kl})
\Bigr]
\end{aligned}
\label{eq:P_ijkl}
\end{equation}

To obtain the trace-less components the following equations are used.
\begin{equation}
\begin{aligned}
C_{ii,jj} &= \frac{1}{3}
 \bigl[4 P_{ii,jj} + P_{jj,jj} + P_{kk,kk}\bigr]
 - 4 P_{ij,ij} - 4 P_{ik,ik} + 2 P_{jk,jk}, \\
C_{ij,ij} &= \frac{1}{3}
 \bigl[-2 P_{ii,ii} - 2 P_{jj,jj} + P_{kk,kk}\bigr]
 + 5 P_{ij,ij} - P_{ik,ik} - P_{jk,jk}, \\
C_{ii,jk} &= \frac{1}{2}
 \bigl[2 P_{ii,jk} - P_{kk,jk} - P_{jj,jk}\bigr], \\
C_{jk,jk} &= \frac{3}{4} P_{jk,jk}
\end{aligned}
\label{eq:C_traceless}
\end{equation}

This leads to 145 unique energy computations for a given basis set. 

\section{Convergence tests}

Convergence tests were performed to find the values for the following C–Pol parameters (using GPAW and Pyscf). All tests were perform on a $H_{2}O_{2}$ monomer oriented as in the figure give in the $C_2$ symmetry section in the SI.

\subsection{Grid-spacing (GPAW)}

\begin{figure}[!ht]
    \centering
\includegraphics[width=0.8\paperwidth]{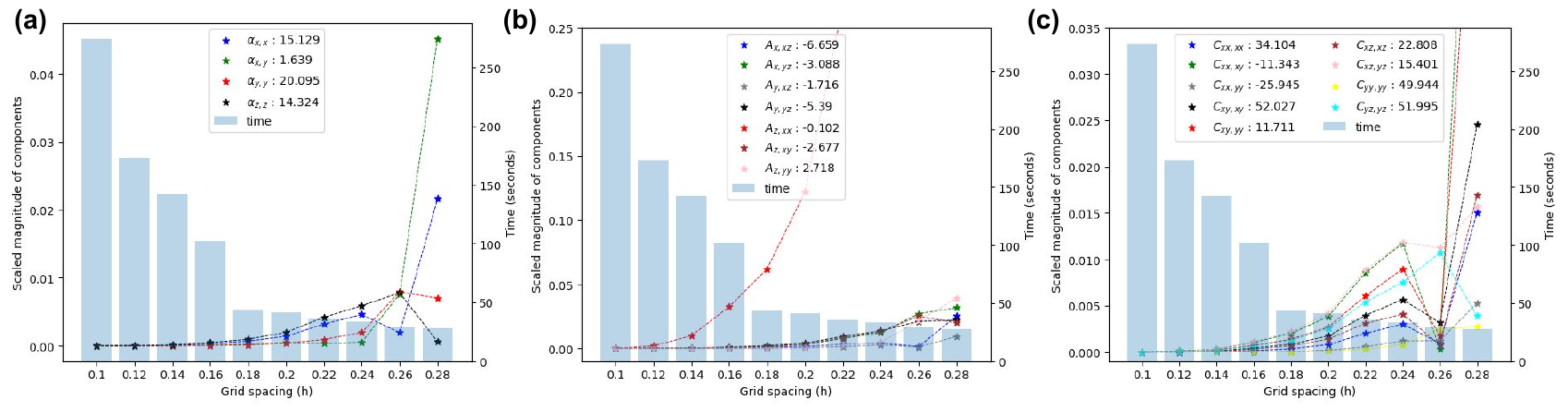}
        \captionof{figure}{The (a) dipole-dipole (b) dipole-quadrupole (c) quadrupole-quadrupole tensor element variation with respect to grid spacing. The tensor elements are base-line shifted and subsequently normalized with respect to its value at a grid spacing of 0.1.}
    \label{fig:convgrid}
\end{figure}

The grid spacing was varied from 0.1 (high resolution) to 0.28 (low resolution.). As seen in figure \ref{fig:convgrid}, the value of 0.18 can be chosen, considering the trade-off between computational time and convergence accuracy.

\subsection{Vacuum (GPAW)}

\begin{figure}[!ht]
    \centering
        \includegraphics[width=0.8\paperwidth]{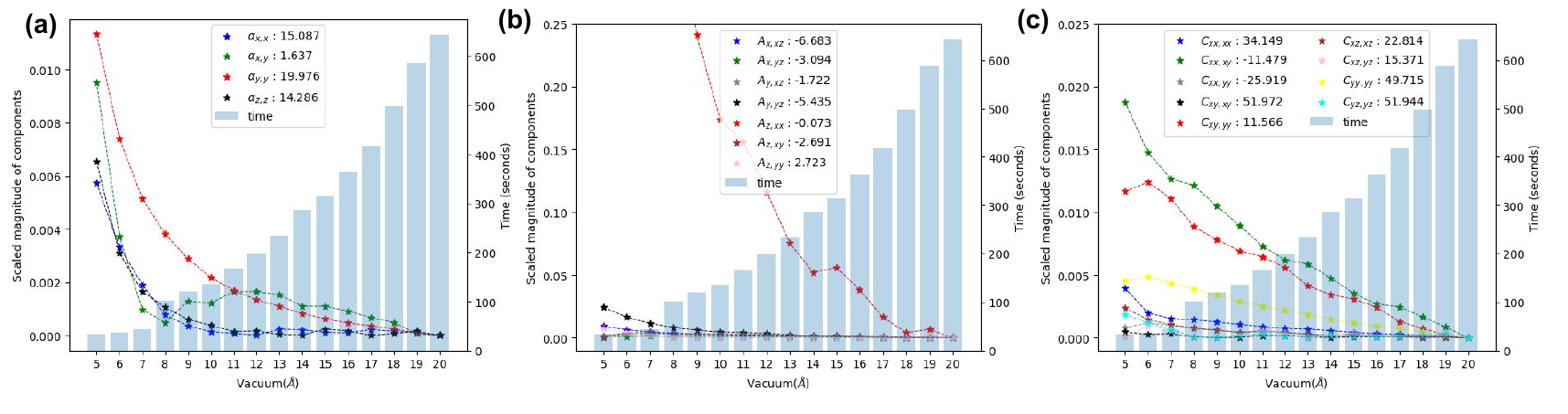}
        \captionof{figure}{The (a) dipole-dipole (b) dipole-quadrupole (c) quadrupole-quadrupole tensor element variation with respect to vacuum. The tensor elements are base-line shifted and subsequently normalized with respect to its value at a vacuum of 20\AA.}
    \label{fig:convvac}
\end{figure}

The vacuum was varied from 5 \AA \: to 20 \AA. As seen in figure \ref{fig:convvac}, the value of 7 \AA \:can be chosen, considering the trade-off between computational time and convergence accuracy.

\subsection{Relative perturbation strength (GPAW)}

The relative perturbation strength (the factor by which the intial potential field (0.009 a.u) and potential field gradient (0.0003 a.u) was scaled) was varied from  $10^{0}$ to $10^{-9}$. As seen in figure \ref{fig:convvac}, the tensor elements converge in the range of $10^{-2}$ to $10^{-7}$.

\begin{figure}[!ht]
    \centering
        \includegraphics[width=0.8\paperwidth]{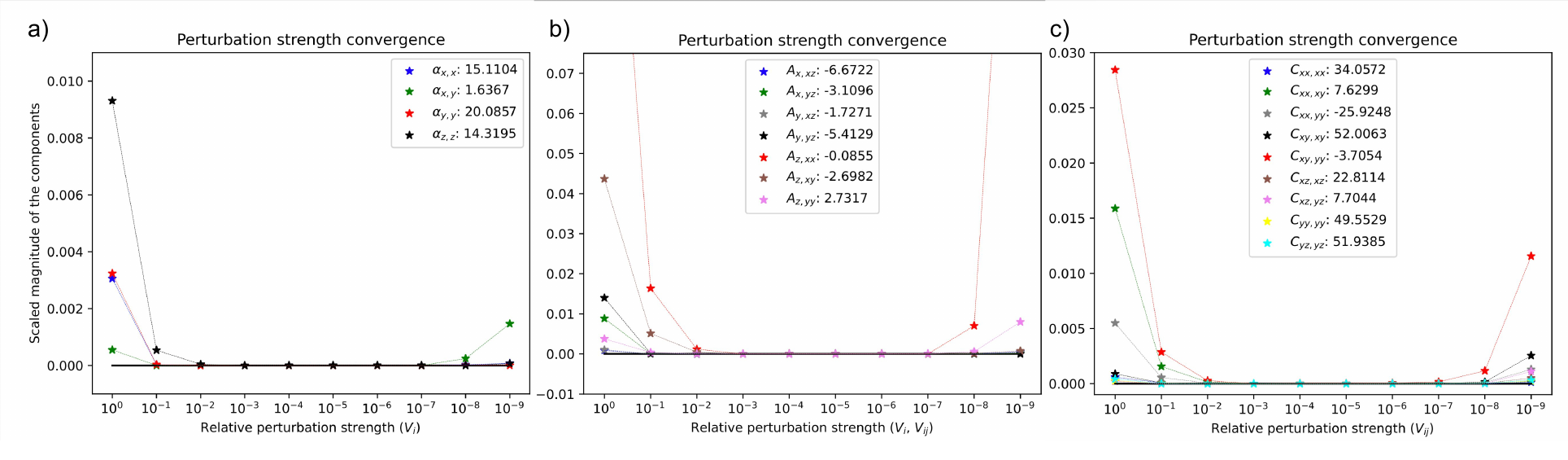}
        \captionof{figure}{The (a) dipole-dipole (b) dipole-quadrupole (c) quadrupole-quadrupole tensor element variation with respect to magnitude of the perturbation. The tensor elements are base-line shifted and subsequently normalized with respect to its value at a relative perturbation strength of $10^{-3}$ (this was chosen after observing the values platuea between $10^{-2}$ and $10^{-7}$).}
    \label{fig:convpert}
\end{figure}

\subsection{Relative perturbation strength (PySCF)}

The relative perturbation strength (the factor by which the intial potential field (0.009 a.u) and potential field gradient (0.0003 a.u) was scaled) was varied from  $10^{0}$ to $10^{-9}$. As seen in figure \ref{fig:convvac}, the tensor elements converge in the range of $10^{-2}$ to $10^{-6}$.

\begin{figure}[!ht]
    \centering
        \includegraphics[width=0.8\paperwidth]{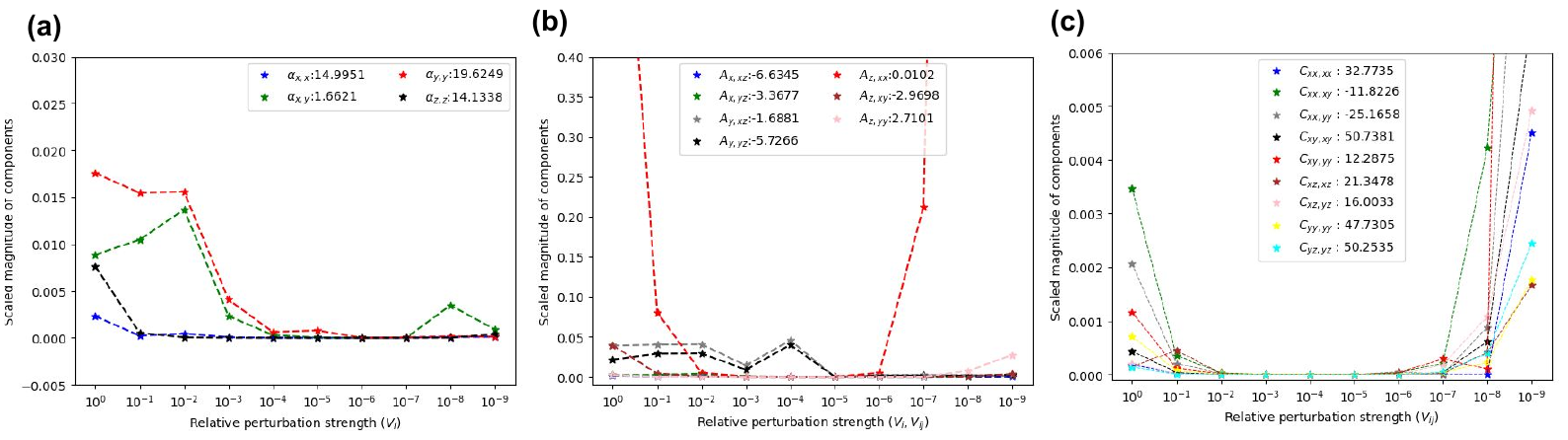}
        \captionof{figure}{The (a) dipole-dipole (b) dipole-quadrupole (c) quadrupole-quadrupole tensor element variation with respect to magnitude of the perturbation. The tensor elements are base-line shifted and subsequently normalized with respect to its value at a relative perturbation strength of $10^{-6}$ (for $\alpha$) and $10^{-4}$ (this was chosen after observing the values platuea between $10^{-2}$ and $10^{-6}$).}
    \label{fig:convpertpyscf}
\end{figure}

\section{Polarizabilites calculated for each point group}

\begin{figure*}[!th]
    \centering
    \includegraphics[width=0.99\textwidth]{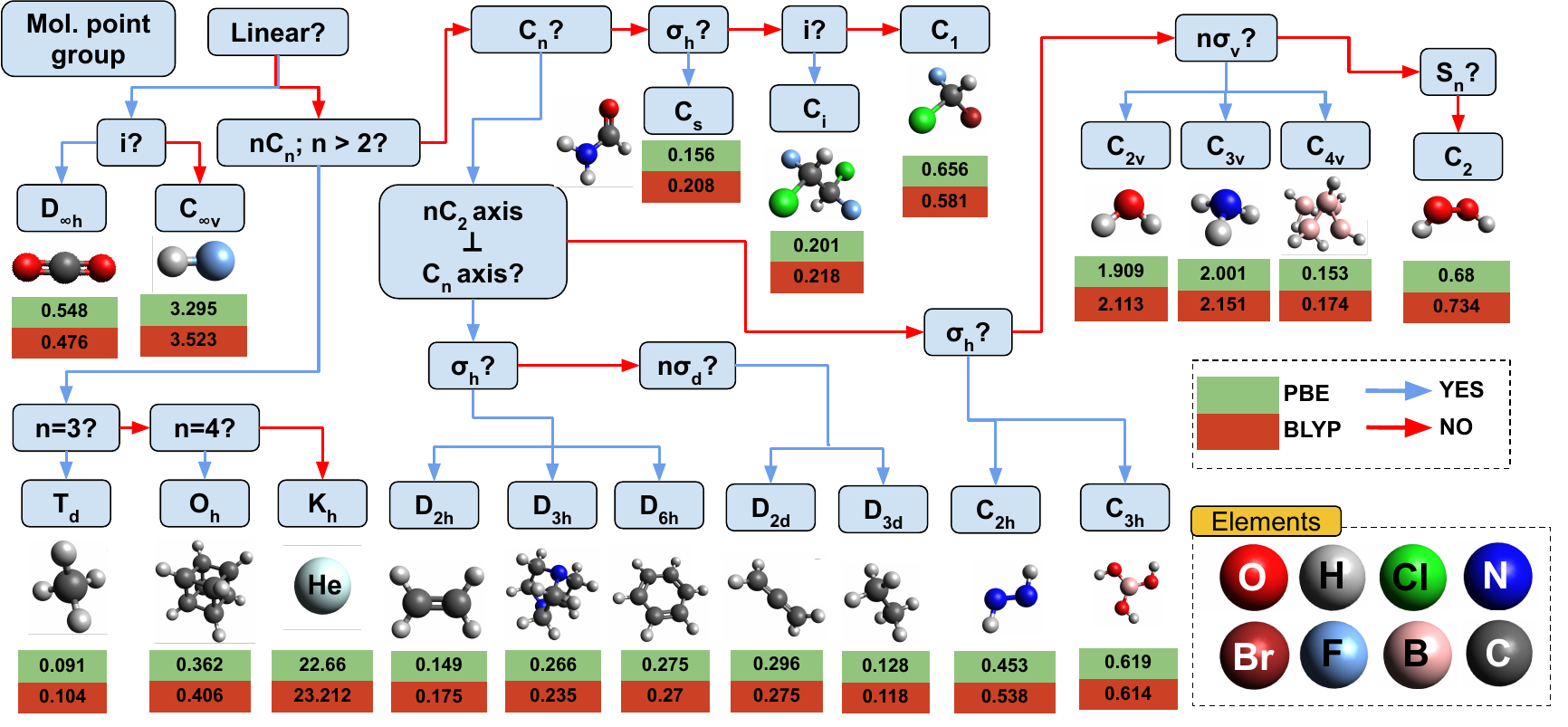} 
    \caption{Molecular point groups studied in this work. For each point group a representative molecule was selected and the polarizability tensors calculated. From left to right, top to bottom : CO$_2$, HF, COHNH$_2$, CH$_2$F$_2$Cl$_2$, CHFClBr, H$_2$O, NH$_3$, B$_5$H$_9$, H$_2$O$_2$, CH$_4$, C$_6$H$_6$, He, C$_2$H$_4$, C$_6$H$_{12}$N$_2$, C$_6$H$_6$, C$_3$H$_4$, C$_2$H$_6$, N$_2$H$_2$ and BO$_3$H$_3$.
    The maximum relative discrepancy between the polarizabilities calculated via G16 and GPAW are shown for PBE (green) and BLYP (orange).}
    \label{fig:pointgroup}
\end{figure*}

The dipole-dipole ($\alpha$), dipole-quadrupole ($A$) and quadrupole-quadrupole ($C$) polarizabilies are calculated for one molecule from each point group using GPAW and G16. This has been done for both PBE and BLYP functionals. In order to use a single metric to compare the G16 and GPAW polarizability tensors, we calculate the max weighted relative discrepancy ($X_{max}$) between GPAW and G16 by multiplying the relative discrepancy observed in a component ($X$) with the fraction of the component's contribution to the sum of all the components as given below:

\begin{equation*}
    X_{max} = Max(\frac{\Delta.|X_{GPAW}|}{\Sigma |X_{GPAW}|})
\end{equation*}

where the relative discrepancy $\Delta$ is given by:

\begin{equation*}
    \Delta = \frac{| X_{GPAW} - X_{G16}|.100}{|X_{G16}|}
\end{equation*}
The maximum of the weighted relative errors for all the tensor elements are calculated and presented in Figure~\ref{fig:pointgroup}  for each molecule using BLYP (red) and PBE (green). 

In general there is a good agreement (maximum relative discrepancy of ~3\% percent or less) between the CBS extrapolated values calculated with G16 and the zero-perturbation extrapolated values calculated with GPAW. Largest discrepancies are observed for components which are numerically small compared to components from the same rank tensor, hence they do not contribute significantly to the electrostatic potential.

\newpage

\newpage
\subsection*{C$_{2v}$-$H_{2}O$}


\begin{minipage}{\linewidth}
    \centering
        \includegraphics[width=0.3\paperwidth]{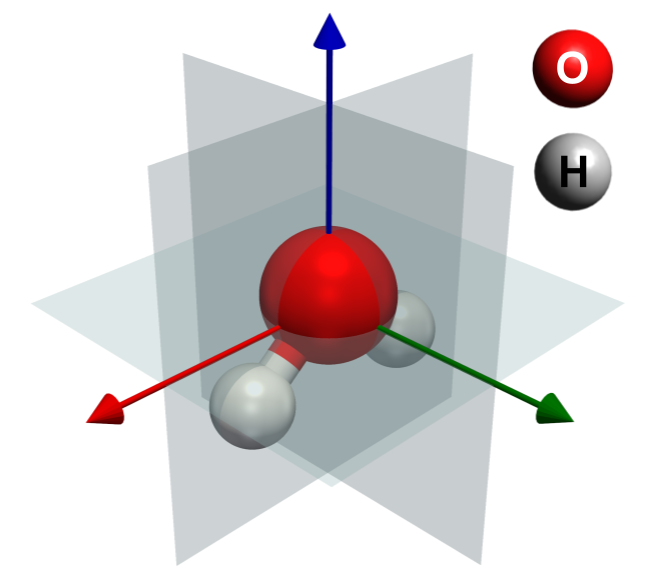}
        \captionof*{figure}{Water ($H_{2}O$)}

\end{minipage}

\begin{table}[!ht]
    \centering

    \rowcolors{2}{gray!25}{white}
\renewcommand{\arraystretch}{1.5} 

\begin{tabular}{|l|r|r|r|r|r|r|}
        \rowcolor{blue!30}
\hline
\textbf{Tensor} &  \multicolumn{3}{c|}{\textbf{PBE}}  & \multicolumn{3}{c|}{\textbf{BLYP}}  \\
\cline{2-7}
        \rowcolor{blue!30}

\textbf{Components} &  \textbf{GPAW} &  \textbf{G16} & \textbf{$\Delta$(\%)} & \textbf{GPAW} &  \textbf{G16}  & \textbf{$\Delta$(\%)} \\
\hline

$\alpha_{x,x}$ & 10.89 & 10.807 & 0.768 & 11.078 & 10.985 & 0.847 \\ 
$\alpha_{y,y}$ & 10.72 & 10.569 & 1.429 & 10.893 & 10.728 & 1.538 \\ 
$\alpha_{z,z}$ & 10.742 & 10.637 & 0.987 & 10.913 & 10.801 & 1.037 \\ 
$A_{x,xz}$ & -7.801 & -7.83 & 0.37 & -8.093 & -8.101 & 0.099 \\ 
$A_{y,yz}$ & -3.151 & -2.989 & 5.42 & -3.206 & -3.049 & 5.149 \\ 
$A_{z,xx}$ & -0.988 & -1.203 & 17.872 & -1.114 & -1.342 & 16.99 \\ 
$A_{z,yy}$ & 4.423 & 4.346 & 1.772 & 4.594 & 4.52 & 1.637 \\ 
$C_{xx,xx}$ & 18.347 & 17.259 & 6.304 & 19.093 & 17.887 & 6.742 \\ 
$C_{xx,yy}$ & -10.37 & -9.587 & 8.167 & -10.848 & -9.952 & 9.003 \\ 
$C_{xy,xy}$ & 14.582 & 13.774 & 5.866 & 15.224 & 14.308 & 6.402 \\ 
$C_{xz,xz}$ & 16.226 & 15.681 & 3.476 & 16.923 & 16.288 & 3.899 \\ 
$C_{yy,yy}$ & 20.891 & 19.151 & 9.086 & 21.859 & 19.881 & 9.949 \\ 
$C_{yz,yz}$ & 14.909 & 13.887 & 7.359 & 15.576 & 14.439 & 7.875 \\

\hline
\end{tabular}

    \caption{The non-zero components of the irreducible representation of the polarizability tensors. The $\alpha$, $A$ and $C$ tensors have 3, 4, and 6 unique elements respectively for the $C_{2v}$.\cite{buckingham}}
\end{table}

The non-zero components of the reducible tensor representation are as follows:
\begin{align*}
    A_{z,zz} &= -(A_{z,xx}+A_{z,yy})\\
    C_{xx,zz} &= -(C_{xx,xx}+C_{xx,yy})\\
    C_{yy,zz} &= -(C_{xx,yy}+C_{yy,yy})\\
    C_{zz,zz} &= -(C_{xx,zz}+C_{yy,zz})\\
\end{align*}

\newpage

\newpage
\subsection*{$D_{6h}-$ $C_{6}H_{6}$}


\begin{minipage}{\linewidth}
    \centering
        \includegraphics[width=0.4\paperwidth]{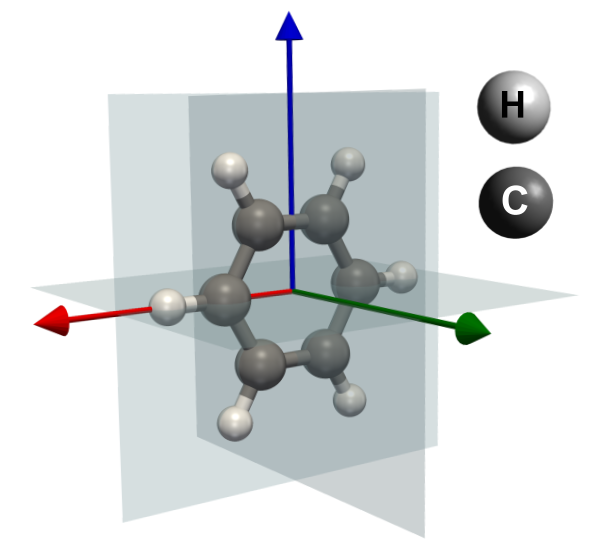}
        \captionof*{figure}{Benzene ($C_{6}H_{6}$)}

\end{minipage}

\begin{table}[!ht]
    \centering

    \rowcolors{2}{gray!25}{white}
\renewcommand{\arraystretch}{1.5} 

\begin{tabular}{|l|r|r|r|r|r|r|}
        \rowcolor{blue!30}
\hline
\textbf{Tensor} &  \multicolumn{3}{c|}{\textbf{PBE}}  & \multicolumn{3}{c|}{\textbf{BLYP}}  \\
\cline{2-7}
        \rowcolor{blue!30}

\textbf{Components} &  \textbf{GPAW} &  \textbf{G16} & \textbf{$\Delta$(\%)} & \textbf{GPAW} &  \textbf{G16}  & \textbf{$\Delta$(\%)} \\
\hline

$\alpha_{x,x}$ & 85.307 & 84.024 & 1.527 & 86.053 & 84.777 & 1.505 \\ 
$\alpha_{y,y}$ & 45.231 & 45.261 & 0.066 & 45.967 & 45.953 & 0.03 \\ 
$C_{xx,xx}$ & 667.411 & 663.641 & 0.568 & 677.093 & 673.318 & 0.561 \\ 
$C_{xy,xy}$ & 293.518 & 293.805 & 0.098 & 302.171 & 302.617 & 0.147 \\ 
$C_{yy,yy}$ & 284.898 & 288.406 & 1.216 & 291.872 & 295.276 & 1.153 \\ 

\hline
\end{tabular}
    \caption{The non-zero components of the irreducible representation of the polarizability tensors. The $\alpha$, $A$ and $C$ tensors have 2, 0, and 3 unique elements respectively for the $D_{6h}$.\cite{buckingham}}
\end{table}

The non-zero components of the reducible tensor representation are as follows \cite{Jahn1937}:
\begin{align*}
    \alpha_{zz} &= \alpha_{xx}\\
    C_{zz,zz} &= C_{xx,xx}\\
    C_{yy,zz} &= C_{xx,yy} = \frac{-C_{yy,yy}}{2}\\
    C_{yz,yz} &= C_{xy,xy}\\
    C_{xz,xz} &= \frac{C_{zz,zz}-C_{xx,zz}}{2}\\
    C_{xx,zz} &= -(C_{xx,xx}+C_{xx,yy})\\
\end{align*}

\newpage
\subsection*{$D_{2d}-$ $C_{3}H_{4}$}


\begin{minipage}{\linewidth}
    \centering
        \includegraphics[width=0.4\paperwidth]{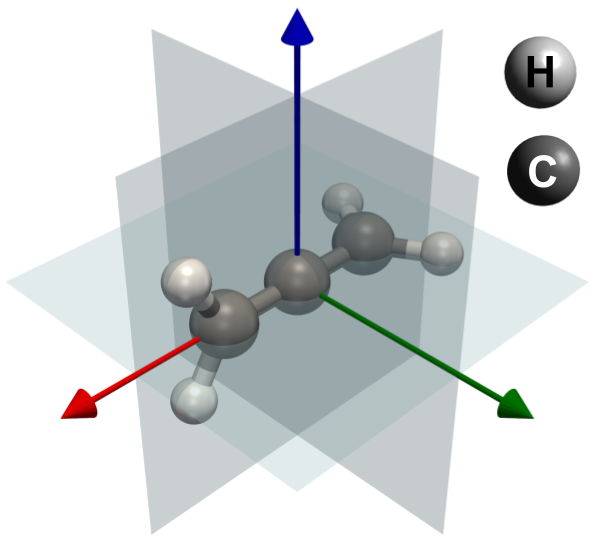}
        \captionof*{figure}{Allene ($C_{3}H_{4}$)}

\end{minipage}

\begin{table}[!ht]
    \centering

    \rowcolors{2}{gray!25}{white}
\renewcommand{\arraystretch}{1.5} 

\begin{tabular}{|l|r|r|r|r|r|r|}
        \rowcolor{blue!30}
\hline
\textbf{Tensor} &  \multicolumn{3}{c|}{\textbf{PBE}}  & \multicolumn{3}{c|}{\textbf{BLYP}}  \\
\cline{2-7}
        \rowcolor{blue!30}

\textbf{Components} &  \textbf{GPAW} &  \textbf{G16} & \textbf{$\Delta$(\%)} & \textbf{GPAW} &  \textbf{G16}  & \textbf{$\Delta$(\%)} \\
\hline

$\alpha_{x,x}$ & 66.103 & 65.009 & 1.683 & 66.718 & 65.646 & 1.633 \\ 
$\alpha_{y,y}$ & 30.852 & 30.908 & 0.181 & 31.181 & 31.224 & 0.138 \\ 
$A_{x,yy}$ & -11.677 & -12.291 & 4.996 & -11.577 & -12.221 & 5.27 \\ 
$A_{y,xy}$ & -12.723 & -12.73 & 0.055 & -12.371 & -12.434 & 0.507 \\ 
$C_{xx,xx}$ & 293.918 & 291.547 & 0.813 & 297.749 & 295.504 & 0.76 \\ 
$C_{xy,xy}$ & 239.241 & 239.061 & 0.075 & 244.011 & 243.862 & 0.061 \\ 
$C_{yy,yy}$ & 135.785 & 135.946 & 0.118 & 139.046 & 139.263 & 0.156 \\ 
$C_{yz,yz}$ & 63.279 & 62.279 & 1.606 & 65.613 & 64.547 & 1.652 \\

\hline
\end{tabular}

    \caption{The non-zero components of the irreducible representation of the polarizability tensors. The $\alpha$, $A$ and $C$ tensors have 2, 2, and 4 unique elements respectively for the $D_{2d}$.\cite{buckingham}}
\end{table}

The non-zero components of the reducible tensor representation are as follows:
\begin{align*}
    \alpha_{zz}  &= \alpha_{yy}\\
    A_{x,zz} &= -A_{x,yy}\\
    A_{z,xz} &= -A_{y,xy}\\
    C_{xx,yy} &= C_{xx,zz} = \frac{-C_{xx,xx}}{2}\\
    C_{xz,xz} &= C_{xy,xy}\\
    C_{yy,zz} &= -(C_{xx,yy}+C_{yy,yy})\\
    C_{zz,zz} &= -(C_{xx,zz}+C_{yy,zz})\\
\end{align*}

\newpage
\subsection*{$D_{2h}-$ $C_{2}H_{4}$}


\begin{minipage}{\linewidth}
    \centering
        \includegraphics[width=0.3\paperwidth]{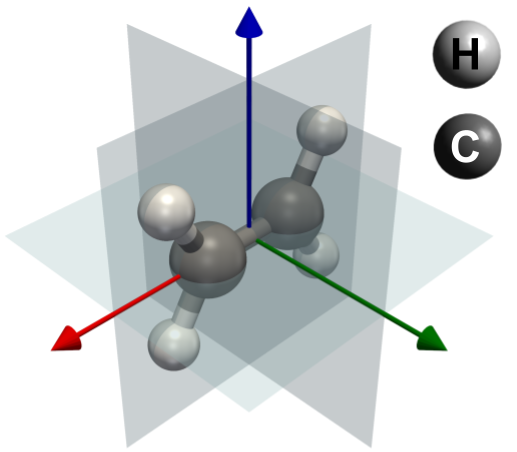}
        \captionof*{figure}{Ethene ($C_{2}H_{4}$)}

\end{minipage}

\begin{table}[!ht]
    \centering

    \rowcolors{2}{gray!25}{white}
\renewcommand{\arraystretch}{1.5} 

\begin{tabular}{|l|r|r|r|r|r|r|}
        \rowcolor{blue!30}
\hline
\textbf{Tensor} &  \multicolumn{3}{c|}{\textbf{PBE}}  & \multicolumn{3}{c|}{\textbf{BLYP}}  \\
\cline{2-7}
        \rowcolor{blue!30}

\textbf{Components} &  \textbf{GPAW} &  \textbf{G16} & \textbf{$\Delta$(\%)} & \textbf{GPAW} &  \textbf{G16}  & \textbf{$\Delta$(\%)} \\
\hline

$\alpha_{x,x}$ & 36.742 & 36.484 & 0.707 & 37.051 & 36.805 & 0.668 \\ 
$\alpha_{y,y}$ & 23.311 & 23.242 & 0.297 & 23.72 & 23.63 & 0.381 \\ 
$\alpha_{z,z}$ & 26.601 & 26.51 & 0.343 & 26.712 & 26.617 & 0.357 \\ 
$C_{xx,xx}$ & 117.1 & 117.027 & 0.062 & 118.851 & 118.658 & 0.163 \\ 
$C_{xx,yy}$ & -60.157 & -60.412 & 0.422 & -61.88 & -62.049 & 0.272 \\ 
$C_{xy,xy}$ & 84.969 & 84.941 & 0.033 & 87.232 & 87.104 & 0.147 \\ 
$C_{xz,xz}$ & 132.568 & 132.372 & 0.148 & 134.571 & 134.36 & 0.157 \\ 
$C_{yy,yy}$ & 83.142 & 82.551 & 0.716 & 86.335 & 85.506 & 0.97 \\ 
$C_{yz,yz}$ & 53.132 & 52.399 & 1.399 & 54.915 & 54.037 & 1.625 \\

\hline
\end{tabular}

    \caption{The non-zero components of the irreducible representation of the polarizability tensors. The $\alpha$, $A$ and $C$ tensors have 3, 0, and 6 unique elements respectively for the $D_{2h}$.\cite{buckingham}}
\end{table}

The non-zero components of the reducible tensor representation are as follows:
\begin{align*}
    C_{xx,zz} &= -(C_{xx,xx}+C_{xx,yy})\\
    C_{yy,zz} &= -(C_{xx,yy}+C_{yy,yy})\\
    C_{zz,zz} &= -(C_{xx,zz}+C_{yy,zz})\\
\end{align*}

\newpage
\subsection*{$C_{2}-$ $H_{2}O_{2}$}


\begin{minipage}{\linewidth}
    \centering
        \includegraphics[width=0.4\paperwidth]{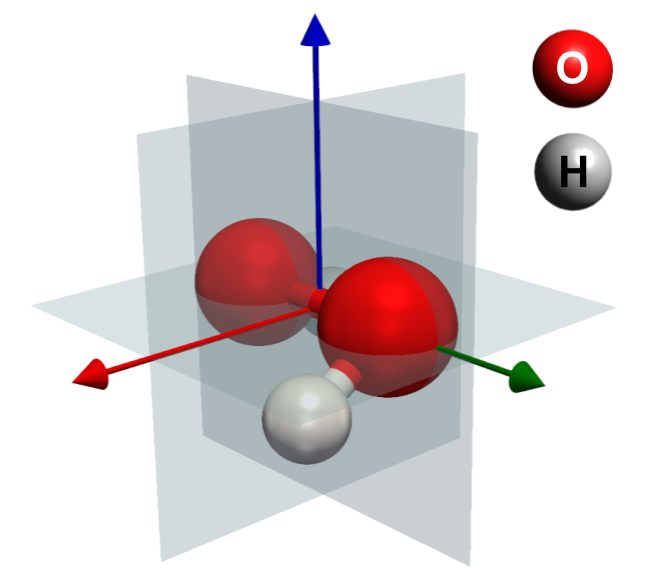}
        \captionof*{figure}{Hydrogen peroxide ($H_{2}O_{2}$)}

\end{minipage}
\begin{table}[!ht]
    \centering

    \rowcolors{2}{gray!25}{white}
\renewcommand{\arraystretch}{1.5} 

\begin{tabular}{|l|r|r|r|r|r|r|}
        \rowcolor{blue!30}
\hline
\textbf{Tensor} &  \multicolumn{3}{c|}{\textbf{PBE}}  & \multicolumn{3}{c|}{\textbf{BLYP}}  \\
\cline{2-7}
        \rowcolor{blue!30}

\textbf{Components} &  \textbf{GPAW} &  \textbf{G16} & \textbf{$\Delta$(\%)} & \textbf{GPAW} &  \textbf{G16}  & \textbf{$\Delta$(\%)} \\
\hline

$\alpha_{x,x}$ & 15.114 & 15.038 & 0.505 & 15.416 & 15.326 & 0.587 \\ 
$\alpha_{x,y}$ & 1.639 & 1.625 & 0.862 & 1.694 & 1.675 & 1.134 \\ 
$\alpha_{y,y}$ & 20.076 & 19.8 & 1.394 & 20.929 & 20.622 & 1.489 \\ 
$\alpha_{z,z}$ & 14.306 & 14.191 & 0.81 & 14.501 & 14.373 & 0.891 \\ 
$A_{x,xz}$ & -6.652 & -6.552 & 1.526 & -6.775 & -6.664 & 1.666 \\ 
$A_{x,yz}$ & -3.092 & -3.165 & 2.306 & -2.982 & -3.062 & 2.613 \\ 
$A_{y,xz}$ & -1.714 & -1.673 & 2.451 & -1.61 & -1.565 & 2.875 \\ 
$A_{y,yz}$ & -5.368 & -5.185 & 3.529 & -5.183 & -5.011 & 3.432 \\ 
$A_{z,xx}$ & -0.111 & -0.313 & 64.537 & -0.291 & -0.466 & 37.554 \\ 
$A_{z,xy}$ & -2.682 & -2.793 & 3.974 & -2.595 & -2.704 & 4.031 \\ 
$A_{z,yy}$ & 2.717 & 2.616 & 3.861 & 2.789 & 2.691 & 3.642 \\ 
$C_{xx,xx}$ & 34.096 & 33.334 & 2.286 & 35.73 & 34.872 & 2.46 \\ 
$C_{xx,xy}$ & 7.687 & 7.841 & 1.964 & 8.248 & 8.399 & 1.798 \\ 
$C_{xx,yy}$ & -25.944 & -25.484 & 1.805 & -27.284 & -26.755 & 1.977 \\ 
$C_{xy,xy}$ & 51.988 & 51.155 & 1.628 & 54.038 & 53.058 & 1.847 \\ 
$C_{xy,yy}$ & -3.72 & -4.073 & 8.667 & -3.96 & -4.322 & 8.376 \\ 
$C_{xz,xz}$ & 22.789 & 21.872 & 4.193 & 23.58 & 22.546 & 4.586 \\ 
$C_{xz,yz}$ & 7.681 & 7.704 & 0.299 & 7.752 & 7.769 & 0.219 \\ 
$C_{yy,yy}$ & 49.547 & 48.029 & 3.161 & 51.959 & 50.255 & 3.391 \\ 
$C_{yz,yz}$ & 51.908 & 50.946 & 1.888 & 53.811 & 52.642 & 2.221 \\

\hline
\end{tabular}

    \caption{The non-zero components of the irreducible representation of the polarizability tensors. The $\alpha$, $A$ and $C$ tensors have 4, 7, and 9 unique elements respectively for the $C_{2}$.}
\end{table}

The non-zero components of the reducible tensor representation are as follows:
\begin{align*}
    \alpha_{zz} &= \alpha_{yy}\\
    A_{z,zz} &= -(A_{z,xx} + A_{z,yy})\\
    C_{xx,zz} &= -(C_{xx,xx}+C_{xx,yy})\\
    C_{xy,zz} &= -(C_{xy,xx}+C_{xy,yy})\\
    C_{yy,zz} &= -(C_{xx,yy}+C_{yy,yy})\\
    C_{zz,zz} &= -(C_{xx,zz}+C_{yy,zz})\\
\end{align*}

\clearpage

\newpage
\subsection*{$D_{\infty h}$--CO$_2$}


\begin{minipage}{\linewidth}
    \centering
        \includegraphics[width=0.3\paperwidth]{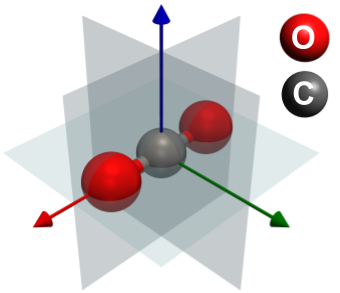}
        \captionof*{figure}{Carbon dioxide ($CO_2$)}

\end{minipage}

\begin{table}[!ht]
    \centering

    \rowcolors{2}{gray!25}{white}
\renewcommand{\arraystretch}{1.5} 

\begin{tabular}{|l|r|r|r|r|r|r|}
        \rowcolor{blue!30}
\hline
\textbf{Tensor} &  \multicolumn{3}{c|}{\textbf{PBE}}  & \multicolumn{3}{c|}{\textbf{BLYP}}  \\
\cline{2-7}
        \rowcolor{blue!30}

\textbf{Components} &  \textbf{GPAW} &  \textbf{G16} & \textbf{$\Delta$(\%)} & \textbf{GPAW} &  \textbf{G16}  & \textbf{$\Delta$(\%)} \\
\hline

$\alpha_{x,x}$ & 27.446 & 27.074 & 1.374 & 27.754 & 27.374 & 1.388 \\ 
$\alpha_{y,y}$ & 13.472 & 13.477 & 0.037 & 13.629 & 13.628 & 0.007 \\ 
$C_{xx,xx}$ & 88.315 & 87.098 & 1.397 & 89.971 & 88.889 & 1.217 \\ 
$C_{xy,xy}$ & 58.326 & 58.428 & 0.175 & 59.953 & 59.853 & 0.167 \\ 
$C_{yy,yy}$ & 37.501 & 36.595 & 2.476 & 38.5 & 37.474 & 2.738 \\

\hline
\end{tabular}

    \caption{The non-zero components of the irreducible representation of the polarizability tensors. The $\alpha$, $A$ and $C$ tensors have 2, 0, and 3 unique elements respectively for the $D_{\infty h}$.\cite{buckingham}}
\end{table}

The non-zero components of the reducible tensor representation are as follows\cite{Jahn1937}:
\begin{align*}
    \alpha_{zz} &= \alpha_{yy}\\
    C_{xx,yy} &= C_{xx,zz} = \frac{-C_{xx,xx}}{2}\\
    C_{xz,xz} &= C_{xy,xy}\\
    C_{yy,zz} &= -(C_{xx,yy}+C_{yy,yy})\\
    C_{zz,zz} &= C_{yy,yy}\\
    C_{yz,yz} &= \frac{C_{yy,yy} - C_{yy,zz}}{2}
\end{align*}

\newpage
\subsection*{$C_{\infty v}-$ $HF$}


\begin{minipage}{\linewidth}
    \centering
        \includegraphics[width=0.35\paperwidth]{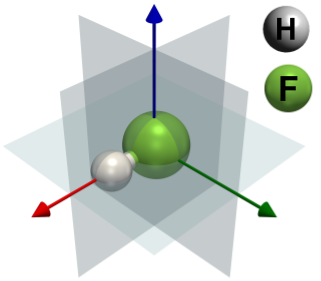}
        \captionof*{figure}{Hydrogen flouride ($HF$)}

\end{minipage}

\begin{table}[!ht]
    \centering

    \rowcolors{2}{gray!25}{white}
\renewcommand{\arraystretch}{1.5} 

\begin{tabular}{|l|r|r|r|r|r|r|}
        \rowcolor{blue!30}
\hline
\textbf{Tensor} &  \multicolumn{3}{c|}{\textbf{PBE}}  & \multicolumn{3}{c|}{\textbf{BLYP}}  \\
\cline{2-7}
        \rowcolor{blue!30}

\textbf{Components} &  \textbf{GPAW} &  \textbf{G16} & \textbf{$\Delta$(\%)} & \textbf{GPAW} &  \textbf{G16}  & \textbf{$\Delta$(\%)} \\
\hline
$\alpha_{x,x}$ & 7.107 & 7.026 & 1.153 & 7.218 & 7.14 & 1.092 \\ 
$\alpha_{y,y}$ & 6.01 & 5.889 & 2.055 & 6.085 & 5.961 & 2.08 \\ 
$A_{x,xx}$ & 5.278 & 5.126 & 2.965 & 5.451 & 5.311 & 2.636 \\ 
$A_{y,xy}$ & 1.636 & 1.557 & 5.074 & 1.662 & 1.598 & 4.005 \\ 
$C_{xx,xx}$ & 10.824 & 10.292 & 5.169 & 11.237 & 10.662 & 5.393 \\ 
$C_{xy,xy}$ & 6.788 & 6.455 & 5.159 & 7.028 & 6.66 & 5.526 \\ 
$C_{yy,yy}$ & 9.247 & 7.923 & 16.711 & 9.594 & 8.149 & 17.732 \\ 

\hline
\end{tabular}

    \caption{The non-zero components of the irreducible representation of the polarizability tensors. The $\alpha$, $A$ and $C$ tensors have 2, 2, and 3 unique elements respectively for the $C_{\infty v}$.\cite{buckingham}}
\end{table}

The non-zero components of the reducible tensor representation are as follows\cite{Jahn1937}:
\begin{align*}
    \alpha_{zz} &= \alpha_{xx}\\
    A_{x,yy} &= A_{x,zz} = \frac{-A_{x,xx}}{2}\\
    A_{z,xz} &= A_{y,xy}\\
    C_{xx,yy} &= C_{xx,zz} = \frac{-C_{xx,xx}}{2}\\
    C_{xz,xz} &= C_{xy,xy}\\
    C_{yy,zz} &= -(C_{xx,yy}+C_{yy,yy})\\
    C_{zz,zz} &= C_{yy,yy}\\
    C_{yz,yz} &= \frac{C_{yy,yy}-C_{yy,zz}}{2}\\
\end{align*}

\newpage
\subsection*{$C_{s}-$ $CHONH_{2}$}


\begin{minipage}{\linewidth}
    \centering
        \includegraphics[width=0.4\paperwidth]{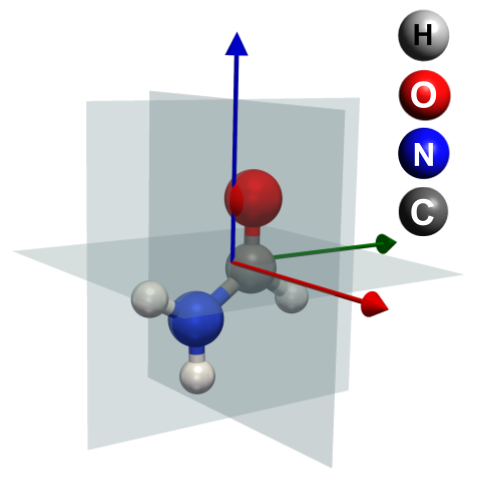}
        \captionof*{figure}{Formamide ($CHONH_{2}$)}

\end{minipage}

\begin{table}[!ht]
    \centering

    \rowcolors{2}{gray!25}{white}
\renewcommand{\arraystretch}{1.5} 

\begin{tabular}{|l|r|r|r|r|r|r|}
        \rowcolor{blue!30}
\hline
\textbf{Tensor} &  \multicolumn{3}{c|}{\textbf{PBE}}  & \multicolumn{3}{c|}{\textbf{BLYP}}  \\
\cline{2-7}
        \rowcolor{blue!30}

\textbf{Components} &  \textbf{GPAW} &  \textbf{G16} & \textbf{$\Delta$(\%)} & \textbf{GPAW} &  \textbf{G16}  & \textbf{$\Delta$(\%)} \\
\hline
$\alpha_{x,x}$ & 21.372 & 21.396 & 0.112 & 21.666 & 21.67 & 0.018 \\ 
$\alpha_{y,y}$ & 32.518 & 32.351 & 0.516 & 32.836 & 32.683 & 0.468 \\ 
$\alpha_{y,z}$ & 3.351 & 3.328 & 0.691 & 3.489 & 3.485 & 0.115 \\ 
$\alpha_{z,z}$ & 37.692 & 37.306 & 1.035 & 38.192 & 37.764 & 1.133 \\ 
$A_{x,xy}$ & -4.831 & -5.071 & 4.733 & -5.366 & -5.641 & 4.875 \\ 
$A_{x,xz}$ & -16.54 & -16.995 & 2.677 & -16.868 & -17.324 & 2.632 \\ 
$A_{y,xx}$ & 2.789 & 2.737 & 1.9 & 2.981 & 2.977 & 0.134 \\ 
$A_{y,yy}$ & -7.706 & -7.65 & 0.732 & -8.077 & -8.168 & 1.114 \\ 
$A_{y,yz}$ & -32.984 & -33.435 & 1.349 & -33.13 & -33.597 & 1.39 \\ 
$A_{z,xx}$ & 15.314 & 15.808 & 3.125 & 15.496 & 15.958 & 2.895 \\ 
$A_{z,yy}$ & 3.479 & 3.522 & 1.221 & 3.656 & 3.714 & 1.562 \\ 
$A_{z,yz}$ & 1.774 & 1.808 & 1.881 & 1.401 & 1.391 & 0.719 \\ 
$C_{xx,xx}$ & 82.773 & 82.463 & 0.376 & 85.371 & 84.796 & 0.678 \\ 
$C_{xx,yy}$ & -28.923 & -28.839 & 0.291 & -29.72 & -29.578 & 0.48 \\ 
$C_{xx,yz}$ & -19.515 & -19.337 & 0.921 & -20.189 & -20.15 & 0.194 \\ 
$C_{xy,xy}$ & 73.711 & 73.048 & 0.908 & 75.816 & 75.11 & 0.94 \\ 
$C_{xy,xz}$ & 24.016 & 24.273 & 1.059 & 25.323 & 25.658 & 1.306 \\ 
$C_{xz,xz}$ & 95.548 & 94.656 & 0.942 & 98.912 & 97.694 & 1.247 \\ 
$C_{yy,yy}$ & 149.27 & 148.446 & 0.555 & 153.21 & 152.211 & 0.656 \\ 
$C_{yy,yz}$ & -3.611 & -3.509 & 2.907 & -3.575 & -3.581 & 0.168 \\ 
$C_{yz,yz}$ & 145.629 & 145.66 & 0.021 & 148.885 & 148.842 & 0.029 \\ 
\hline
\end{tabular}

    \caption{The non-zero components of the irreducible representation of the polarizability tensors. The $\alpha$, $A$ and $C$ tensors have 4, 8, and 9 unique elements respectively for the $C_{s}$.\cite{buckingham}}
\end{table}
The non-zero components of the reducible tensor representation are as follows:
\begin{align*}
    A_{y,zz} &= -(A_{y,xx}+A_{y,yy})\\
    A_{z,zz} &= -(A_{z,xx}+A_{z,yy})\\
    C_{xx,zz} &= -(C_{xx,yy}+C_{xx,xx})\\
    C_{yy,zz} &= -(C_{yy,yy}+C_{xx.yy})\\
    C_{yz,zz} &= -(C_{yy,yz}+C_{xx,yz})\\
    C_{zz,zz} &= -(C_{yy,zz}+C_{xx,zz})
\end{align*}

\clearpage

\newpage
\subsection*{$D_{3d}-$ $C_{2}H_{6}$}


\begin{minipage}{\linewidth}
    \centering
        \includegraphics[width=0.4\paperwidth]{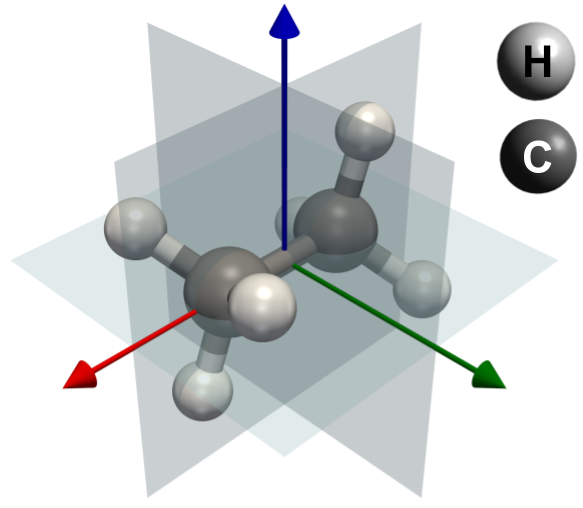}
        \captionof*{figure}{Ethane ($C_{2}H_{6}$)}

\end{minipage}

\begin{table}[!ht]
    \centering

    \rowcolors{2}{gray!25}{white}
\renewcommand{\arraystretch}{1.5} 

\begin{tabular}{|l|r|r|r|r|r|r|}
        \rowcolor{blue!30}
\hline
\textbf{Tensor} &  \multicolumn{3}{c|}{\textbf{PBE}}  & \multicolumn{3}{c|}{\textbf{BLYP}}  \\
\cline{2-7}
        \rowcolor{blue!30}

\textbf{Components} &  \textbf{GPAW} &  \textbf{G16} & \textbf{$\Delta$(\%)} & \textbf{GPAW} &  \textbf{G16}  & \textbf{$\Delta$(\%)} \\
\hline
$\alpha_{x,x}$ & 33.543 & 33.173 & 1.115 & 33.635 & 33.306 & 0.988 \\ 
$\alpha_{y,y}$ & 28.803 & 28.734 & 0.24 & 28.868 & 28.806 & 0.215 \\ 
$C_{xx,xx}$ & 143.329 & 142.72 & 0.427 & 145.323 & 144.877 & 0.308 \\ 
$C_{xy,xy}$ & 149.184 & 148.885 & 0.201 & 151.641 & 151.455 & 0.123 \\ 
$C_{xy,yz}$ & 19.385 & 19.449 & 0.329 & 20.113 & 20.199 & 0.426 \\ 
$C_{yy,yy}$ & 100.226 & 100.709 & 0.48 & 101.962 & 102.538 & 0.562 \\

\hline
\end{tabular}

    \caption{The non-zero components of the irreducible representation of the polarizability tensors. The $\alpha$, $A$ and $C$ tensors have 2, 0, and 4 unique elements respectively for the $D_{3d}$\cite{buckingham}.}
\end{table}
The non-zero components of the reducible tensor representation are as follows:
\begin{align*}
    \alpha_{zz} &= \alpha_{yy}\\
    C_{xx,yy} &= C_{xx,zz} = \frac{-C_{xx,xx}}{2}\\
    C_{xz,xz} &= C_{xy,xy}\\
    C_{xz,yy} &= C_{xy,yz}\\
    C_{xz,zz} &= -C_{xz,yy}\\
    C_{yz,yz} &= \frac{C_{yy,yy}-C_{yy,zz}}{2}\\
    C_{yy,zz} &= -(C_{yy,yy}+C_{xx,yy})\\
    C_{zz,zz} &= C_{yy,yy}\\
\end{align*}

\newpage
\subsection*{$D_{3h}-$ $C_{6}H_{12}N_{2}$}


\begin{minipage}{\linewidth}
    \centering
        \includegraphics[width=0.4\paperwidth]{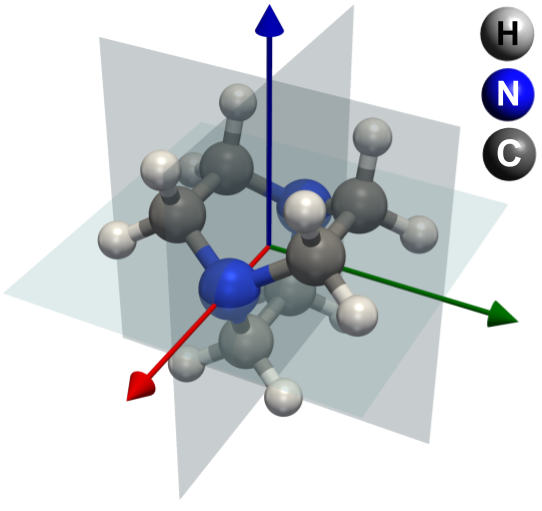}
        \captionof*{figure}{Triethylenediamine ($C_{6}H_{12}N_{2}$)}

\end{minipage}

\begin{table}[!ht]
    \centering

    \rowcolors{2}{gray!25}{white}
\renewcommand{\arraystretch}{1.5} 

\begin{tabular}{|l|r|r|r|r|r|r|}
        \rowcolor{blue!30}
\hline
\textbf{Tensor} &  \multicolumn{3}{c|}{\textbf{PBE}}  & \multicolumn{3}{c|}{\textbf{BLYP}}  \\
\cline{2-7}
        \rowcolor{blue!30}

\textbf{Components} &  \textbf{GPAW} &  \textbf{G16} & \textbf{$\Delta$(\%)} & \textbf{GPAW} &  \textbf{G16}  & \textbf{$\Delta$(\%)} \\
\hline
$\alpha_{x,x}$ & 88.386 & 87.657 & 0.832 & 89.391 & 88.55 & 0.95 \\ 
$\alpha_{y,y}$ & 92.272 & 91.376 & 0.981 & 92.695 & 91.781 & 0.996 \\ 
$A_{y,yz}$ & -5.274 & -5.189 & 1.638 & -3.548 & -3.621 & 2.016 \\ 
$C_{xx,xx}$ & 689.922 & 691.027 & 0.16 & 711.606 & 712.343 & 0.103 \\ 
$C_{xy,xy}$ & 732.14 & 733.944 & 0.246 & 744.173 & 745.245 & 0.144 \\ 
$C_{yy,yy}$ & 846.456 & 853.004 & 0.768 & 861.591 & 867.498 & 0.681 \\

\hline
\end{tabular}

    \caption{The non-zero components of the irreducible representation of the polarizability tensors. The $\alpha$, $A$ and $C$ tensors have 2, 1, and 3 unique elements respectively for the $D_{3h}$.\cite{buckingham}}
\end{table}

The non-zero components of the reducible tensor representation are as follows\cite{Jahn1937}:
\begin{align*}
    \alpha_{zz} &= \alpha_{yy}\\
    A_{z,yy} &= A_{y,yz}\\
    A_{z,zz} &= -A_{z,yy}\\
    C_{xx,yy} &= C_{xx,zz} = \frac{-C_{xx,xx}}{2}\\
    C_{xz,xz} &= C_{xy,xy}\\
    C_{yy,zz} &= -(C_{xx,yy}+C_{yy,yy})\\
    C_{yz,yz} &= \frac{C_{yy,yy}-C_{yy,zz}}{2}\\
    C_{zz,zz} &= C_{yy,yy}
\end{align*}

\newpage
\subsection*{$C_{4v}-$ $B_{5}H_{9}$}


\begin{minipage}{\linewidth}
    \centering
        \includegraphics[width=0.4\paperwidth]{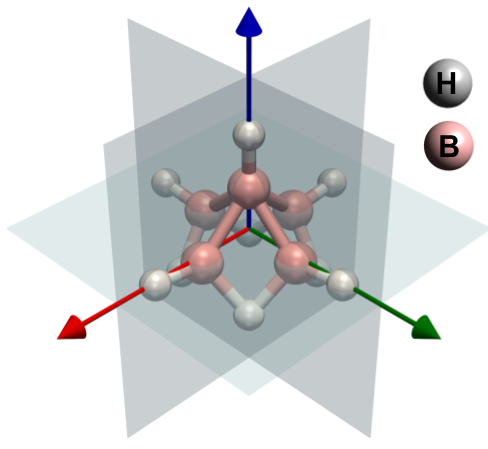}
        \captionof*{figure}{Pentaborane ($B_{5}H_{9}$)}

\end{minipage}

\begin{table}[!ht]
    \centering

    \rowcolors{2}{gray!25}{white}
\renewcommand{\arraystretch}{1.5} 

\begin{tabular}{|l|r|r|r|r|r|r|}
        \rowcolor{blue!30}
\hline
\textbf{Tensor} &  \multicolumn{3}{c|}{\textbf{PBE}}  & \multicolumn{3}{c|}{\textbf{BLYP}}  \\
\cline{2-7}
        \rowcolor{blue!30}

\textbf{Components} &  \textbf{GPAW} &  \textbf{G16} & \textbf{$\Delta$(\%)} & \textbf{GPAW} &  \textbf{G16}  & \textbf{$\Delta$(\%)} \\
\hline

$\alpha_{x,x}$ & 76.546 & 75.582 & 1.275 & 76.23 & 75.273 & 1.271 \\ 
$\alpha_{z,z}$ & 65.328 & 64.839 & 0.754 & 64.952 & 63.948 & 1.57 \\ 
$A_{x,xz}$ & -29.587 & -30.364 & 2.559 & -29.677 & -30.3 & 2.056 \\ 
$A_{z,zz}$ & 34.796 & 34.79 & 0.017 & 34.424 & 34.05 & 1.098 \\ 
$C_{xx,xx}$ & 730.865 & 727.633 & 0.444 & 734.632 & 730.938 & 0.505 \\ 
$C_{xy,xy}$ & 385.862 & 386.104 & 0.063 & 387.954 & 388.533 & 0.149 \\ 
$C_{xz,xz}$ & 341.893 & 342.898 & 0.293 & 343.889 & 345.268 & 0.399 \\ 
$C_{zz,zz}$ & 510.007 & 507.808 & 0.433 & 510.585 & 508.492 & 0.412 \\ 

\hline
\end{tabular}

    \caption{The non-zero components of the irreducible representation of the polarizability tensors. The $\alpha$, $A$ and $C$ tensors have 2, 2, and 4 unique elements respectively for the $C_{4v}$.\cite{buckingham}}
\end{table}

The non-zero components of the reducible tensor representation are as follows:
\begin{align*}
    \alpha_{yy} &= \alpha_{xx}\\
    A_{z,xx} &= A_{z,yy} = \frac{-A_{z,zz}}{2}\\
    A_{y,yz} &= A_{x,xz}\\
    C_{yy,yy} &= C_{xx,xx}\\
    C_{yz,yz} &= C_{xy,xy}\\
    C_{xx,zz} &= C_{yy,zz} = \frac{-C_{zz,zz}}{2}\\
    C_{xx,yy} &= -(C_{xx,xx} + C_{xx,zz})\\
\end{align*}

\newpage
\subsection*{$C_{3v}-$ $NH_{3}$}


\begin{minipage}{\linewidth}
    \centering
        \includegraphics[width=0.35\paperwidth]{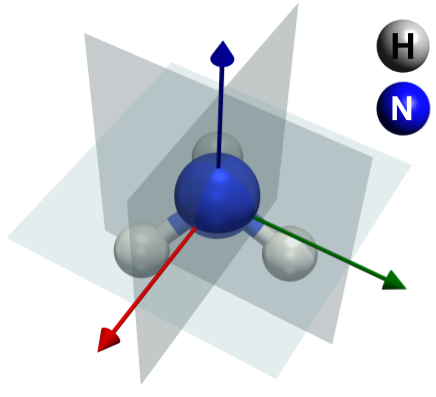}
        \captionof*{figure}{Ammonia ($NH_{3}$)}

\end{minipage}

\begin{table}[!ht]
    \centering

    \rowcolors{2}{gray!25}{white}
\renewcommand{\arraystretch}{1.5} 

\begin{tabular}{|l|r|r|r|r|r|r|}
        \rowcolor{blue!30}
\hline
\textbf{Tensor} &  \multicolumn{3}{c|}{\textbf{PBE}}  & \multicolumn{3}{c|}{\textbf{BLYP}}  \\
\cline{2-7}
        \rowcolor{blue!30}

\textbf{Components} &  \textbf{GPAW} &  \textbf{G16} & \textbf{$\Delta$(\%)} & \textbf{GPAW} &  \textbf{G16}  & \textbf{$\Delta$(\%)} \\
\hline

$\alpha_{x,x}$ & 14.732 & 14.67 & 0.423 & 14.927 & 14.855 & 0.485 \\ 
$\alpha_{z,z}$ & 17.459 & 17.282 & 1.024 & 17.803 & 17.607 & 1.113 \\ 
$A_{x,xz}$ & -5.869 & -5.953 & 1.411 & -6.001 & -6.039 & 0.629 \\ 
$A_{y,xx}$ & -5.848 & -5.897 & 0.831 & -6.089 & -6.156 & 1.088 \\ 
$A_{z,zz}$ & -1.822 & -1.765 & 3.229 & -1.696 & -1.607 & 5.538 \\ 
$C_{xx,xx}$ & 32.344 & 31.379 & 3.075 & 33.651 & 32.464 & 3.656 \\ 
$C_{xx,yy}$ & -13.616 & -13.647 & 0.227 & -13.958 & -13.895 & 0.453 \\ 
$C_{xx,yz}$ & 2.121 & 2.508 & 15.431 & 2.246 & 2.672 & 15.943 \\ 
$C_{xz,xz}$ & 29.619 & 28.903 & 2.477 & 31.18 & 30.3 & 2.904 \\

\hline
\end{tabular}

    \caption{The non-zero components of the irreducible representation of the polarizability tensors. The $\alpha$, $A$ and $C$ tensors have 2, 3, and 4 unique elements respectively for the $C_{3v}$.}
\end{table}

The non-zero components of the reducible tensor representation are as follows:
\begin{align*}
    \alpha_{y,y} &= \alpha_{x,x}\\
    A_{z,xx} &= A_{z,yy} = \frac{-A_{z,zz}}{2}\\
    A_{y,yy} &= -A_{y,xx}\\
    A_{x,xy} &= A_{y,xx}\\
    A_{y,yz} &= A_{x,xz}\\
    C_{yy,yy} &= C_{xx,xx} \\
    C_{yy,zz} &= C_{xx,zz}\\
    C_{yz,yz} &= C_{xz,xz}\\
    C_{yy,yz} &= -C_{xx,yz}\\
    C_{xy,xz} &= C_{xx,yz}\\
    C_{xy,xy} &= \frac{C_{xx,xx}-C_{xx,yy}}{2}\\
    C_{xx,zz} &= -(C_{xx,yy}+C_{xx,xx})\\
    C_{zz,zz} &= -(C_{yy,zz}+C_{xx,zz})\\
\end{align*}

\newpage
\subsection*{$C_{2h}-$$N_{2}H_{2}$}


\begin{minipage}{\linewidth}
    \centering
        \includegraphics[width=0.4\paperwidth]{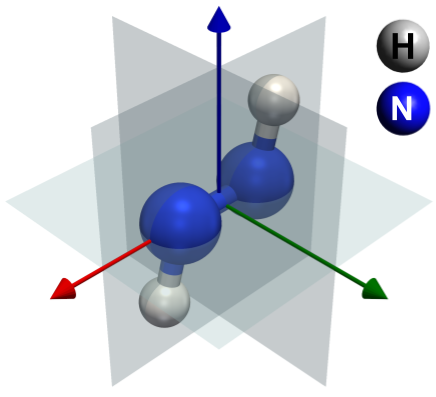}
        \captionof*{figure}{Diazene ($N_{2}H_{2}$)}

\end{minipage}

\begin{table}[!ht]
    \centering

    \rowcolors{2}{gray!25}{white}
\renewcommand{\arraystretch}{1.5} 

\begin{tabular}{|l|r|r|r|r|r|r|}
        \rowcolor{blue!30}
\hline
\textbf{Tensor} &  \multicolumn{3}{c|}{\textbf{PBE}}  & \multicolumn{3}{c|}{\textbf{BLYP}}  \\
\cline{2-7}
        \rowcolor{blue!30}

\textbf{Components} &  \textbf{GPAW} &  \textbf{G16} & \textbf{$\Delta$(\%)} & \textbf{GPAW} &  \textbf{G16}  & \textbf{$\Delta$(\%)} \\
\hline

$\alpha_{x,x}$ & 25.144 & 24.901 & 0.976 & 25.546 & 25.289 & 1.016 \\ 
$\alpha_{x,z}$ & -2.915 & -2.921 & 0.205 & -2.912 & -2.928 & 0.546 \\ 
$\alpha_{y,y}$ & 14.495 & 14.432 & 0.437 & 14.716 & 14.659 & 0.389 \\ 
$\alpha_{z,z}$ & 20.97 & 21.04 & 0.333 & 21.164 & 21.237 & 0.344 \\ 
$C_{xx,xx}$ & 65.844 & 66.15 & 0.463 & 67.625 & 67.915 & 0.427 \\ 
$C_{xx,xz}$ & 9.973 & 11.012 & 9.435 & 10.172 & 11.323 & 10.165 \\ 
$C_{xx,yy}$ & -28.699 & -27.796 & 3.249 & -29.83 & -28.88 & 3.289 \\ 
$C_{xy,xy}$ & 52.895 & 52.322 & 1.095 & 54.598 & 53.945 & 1.21 \\ 
$C_{xy,yz}$ & 5.194 & 4.556 & 14.004 & 5.654 & 4.929 & 14.709 \\ 
$C_{xz,xz}$ & 83.178 & 81.688 & 1.824 & 85.889 & 84.077 & 2.155 \\ 
$C_{xz,yy}$ & 11.023 & 11.148 & 1.121 & 11.159 & 11.325 & 1.466 \\ 
$C_{yy,yy}$ & 45.479 & 44.392 & 2.449 & 47.328 & 46.051 & 2.773 \\ 
$C_{yz,yz}$ & 31.683 & 30.659 & 3.34 & 32.669 & 31.413 & 3.998 \\

\hline
\end{tabular}

    \caption{The non-zero components of the irreducible representation of the polarizability tensors. The $\alpha$, $A$ and $C$ tensors have 4, 0, and 9 unique elements respectively for the $C_{2h}$.\cite{buckingham}}
\end{table}

The non-zero components of the reducible tensor representation are as follows\cite{Jahn1937}:
\begin{align*}
    C_{xx,zz} &= -(C_{xx,xx} + C_{xx,yy})\\
    C_{xz,zz} &= -(C_{xz,yy} + C_{xx,xz})\\
    C_{yy,zz} &= -(C_{yy,yy} + C_{xx,yy})\\
    C_{zz,zz} &= -(C_{xx,zz} + C_{yy,zz})\\
\end{align*}

\clearpage
\newpage
\subsection*{$C_{1}-$$CHFClBr$}


\begin{minipage}{\linewidth}
    \centering
        \includegraphics[width=0.4\paperwidth]{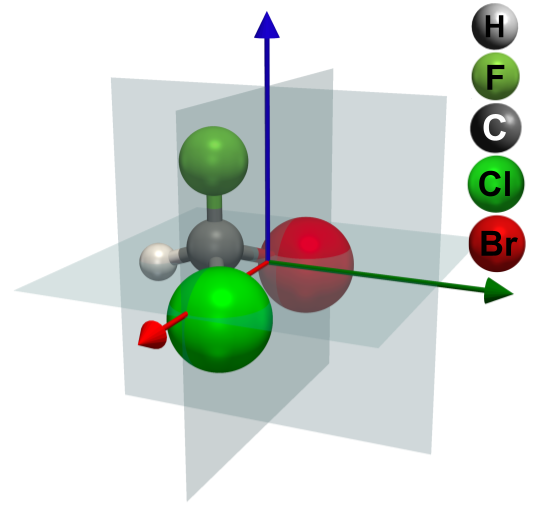}
        \captionof*{figure}{Bromochlorofluoromethane ($CHFClBr$)}

\end{minipage}

\newpage
The non-zero components of the reducible tensor representation are as follows\cite{Jahn1937}:
\begin{align*}
    A_{x,zz} &= -(A_{x,xx} + A_{x,yy})\\
    A_{y,zz} &= -(A_{y,xx} + A_{y,yy})\\
    A_{z,zz} &= -(A_{z,xx} + A_{z,yy})\\
    C_{xx,zz} &= -(C_{xx,xx} + C_{xx,yy})\\
    C_{xy,zz} &= -(C_{xy,yy} + C_{xx,xy})\\
    C_{xz,zz} &= -(C_{xz,yy} + C_{xx,xz})\\
    C_{yy,zz} &= -(C_{yy,yy} + C_{xx,yy})\\
    C_{yz,zz} &= -(C_{yy,yz} + C_{xx,yz})\\
    C_{zz,zz} &= -(C_{xx,zz} + C_{yy,zz})\\
\end{align*}

\begin{table}[!ht]
    \centering

    \rowcolors{2}{gray!25}{white}
\renewcommand{\arraystretch}{1.5} 

\begin{tabular}{|l|r|r|r|r|r|r|}
        \rowcolor{blue!30}
\hline
\textbf{Tensor} &  \multicolumn{3}{c|}{\textbf{PBE}}  & \multicolumn{3}{c|}{\textbf{BLYP}}  \\
\cline{2-7}
        \rowcolor{blue!30}

\textbf{Components} &  \textbf{GPAW} &  \textbf{G16} & \textbf{$\Delta$(\%)} & \textbf{GPAW} &  \textbf{G16}  & \textbf{$\Delta$(\%)} \\
\hline

$\alpha_{x,x}$ & 71.661 & 69.348 & 3.335 & 73.186 & 71.418 & 2.476 \\ 
$\alpha_{x,y}$ & -1.111 & -1.17 & 5.043 & -1.215 & -1.233 & 1.46 \\ 
$\alpha_{x,z}$ & 1.563 & 1.478 & 5.751 & 1.62 & 1.565 & 3.514 \\ 
$\alpha_{y,y}$ & 46.944 & 46.572 & 0.799 & 47.496 & 47.29 & 0.436 \\ 
$\alpha_{y,z}$ & -3.044 & -3.001 & 1.433 & -3.167 & -3.121 & 1.474 \\ 
$\alpha_{z,z}$ & 48.808 & 48.263 & 1.129 & 49.601 & 49.202 & 0.811 \\ 
$A_{x,xx}$ & 79.677 & 77.804 & 2.407 & 83.604 & 81.398 & 2.71 \\ 
$A_{x,xy}$ & 12.213 & 11.799 & 3.509 & 12.882 & 12.754 & 1.004 \\ 
$A_{x,xz}$ & -35.665 & -35.017 & 1.851 & -37.248 & -36.927 & 0.869 \\ 
$A_{x,yy}$ & -39.624 & -38.898 & 1.866 & -41.581 & -40.711 & 2.137 \\ 
$A_{x,yz}$ & -1.713 & -1.565 & 9.457 & -1.503 & -1.704 & 11.796 \\ 
$A_{y,xx}$ & 26.091 & 26.28 & 0.719 & 26.578 & 26.69 & 0.42 \\ 
$A_{y,xy}$ & 31.507 & 32.675 & 3.575 & 32.197 & 33.4 & 3.602 \\ 
$A_{y,xz}$ & -4.22 & -3.983 & 5.95 & -4.198 & -4.231 & 0.78 \\ 
$A_{y,yy}$ & -19.487 & -19.516 & 0.149 & -19.202 & -18.971 & 1.218 \\ 
$A_{y,yz}$ & -8.699 & -8.513 & 2.185 & -8.743 & -8.813 & 0.794 \\ 
$A_{z,xx}$ & -20.851 & -21.284 & 2.034 & -22.053 & -22.25 & 0.885 \\ 
$A_{z,xy}$ & -4.648 & -4.33 & 7.344 & -4.661 & -4.615 & 0.997 \\ 
$A_{z,xz}$ & 33.618 & 34.866 & 3.579 & 34.767 & 35.954 & 3.301 \\ 
$A_{z,yy}$ & 6.096 & 6.927 & 11.997 & 6.512 & 7.49 & 13.057 \\ 
$A_{z,yz}$ & -13.789 & -13.696 & 0.679 & -14.208 & -13.913 & 2.12 \\

\hline
\end{tabular}

    \caption{The non-zero components of the irreducible representation of the polarizability tensors. The $\alpha$ and $A$ tensors have 6 and 15 unique elements respectively for the $C_{1}$.\cite{buckingham}}
\end{table}

\begin{table}[!ht]
    \centering

    \rowcolors{2}{gray!25}{white}
\renewcommand{\arraystretch}{1.5} 

\begin{tabular}{|l|r|r|r|r|r|r|}
        \rowcolor{blue!30}
\hline
\textbf{Tensor} &  \multicolumn{3}{c|}{\textbf{PBE}}  & \multicolumn{3}{c|}{\textbf{BLYP}}  \\
\cline{2-7}
        \rowcolor{blue!30}

\textbf{Components} &  \textbf{GPAW} &  \textbf{G16} & \textbf{$\Delta$(\%)} & \textbf{GPAW} &  \textbf{G16}  & \textbf{$\Delta$(\%)} \\
\hline

$C_{xx,xx}$ & 603.998 & 593.365 & 1.792 & 631.025 & 621.229 & 1.577 \\ 
$C_{xx,xy}$ & 23.278 & 21.352 & 9.02 & 23.21 & 22.865 & 1.509 \\ 
$C_{xx,xz}$ & -29.349 & -28.06 & 4.594 & -30.647 & -30.363 & 0.935 \\ 
$C_{xx,yy}$ & -289.402 & -284.973 & 1.554 & -301.149 & -297.388 & 1.265 \\ 
$C_{xx,yz}$ & 45.756 & 45.599 & 0.344 & 48.094 & 48.174 & 0.166 \\ 
$C_{xy,xy}$ & 339.274 & 335.067 & 1.256 & 352.872 & 350.01 & 0.818 \\ 
$C_{xy,xz}$ & -8.812 & -9.476 & 7.007 & -9.033 & -9.405 & 3.955 \\ 
$C_{xy,yy}$ & -23.942 & -23.247 & 2.99 & -24.51 & -24.195 & 1.302 \\ 
$C_{xy,yz}$ & 10.53 & 9.987 & 5.437 & 11.055 & 10.524 & 5.046 \\ 
$C_{xz,xz}$ & 362.009 & 356.927 & 1.424 & 377.027 & 373.341 & 0.987 \\ 
$C_{xz,yy}$ & -4.242 & -4.087 & 3.793 & -4.314 & -3.821 & 12.902 \\ 
$C_{xz,yz}$ & -13.379 & -13.937 & 4.004 & -14.484 & -14.549 & 0.447 \\ 
$C_{yy,yy}$ & 297.853 & 293.766 & 1.391 & 308.368 & 303.87 & 1.48 \\ 
$C_{yy,yz}$ & -11.008 & -11.158 & 1.344 & -11.804 & -11.987 & 1.527 \\ 
$C_{yz,yz}$ & 141.1 & 137.496 & 2.621 & 146.474 & 141.707 & 3.364 \\

\hline
\end{tabular}

    \caption{The $C$ tensors have 15 unique elements for the $C_{1}$.\cite{buckingham}}
\end{table}

\clearpage
\newpage

\subsection*{$C_{3h}-$ $B(OH)_{3}$}


\begin{minipage}{\linewidth}
    \centering
        \includegraphics[width=0.4\paperwidth]{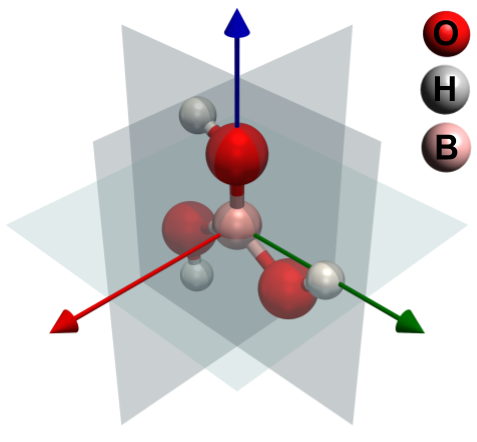}
        \captionof*{figure}{Boric acid ($B(OH)_{3}$)}

\end{minipage}

\begin{table}[!ht]
    \centering

    \rowcolors{2}{gray!25}{white}
\renewcommand{\arraystretch}{1.5} 

\begin{tabular}{|l|r|r|r|r|r|r|}
        \rowcolor{blue!30}
\hline
\textbf{Tensor} &  \multicolumn{3}{c|}{\textbf{PBE}}  & \multicolumn{3}{c|}{\textbf{BLYP}}  \\
\cline{2-7}
        \rowcolor{blue!30}

\textbf{Components} &  \textbf{GPAW} &  \textbf{G16} & \textbf{$\Delta$(\%)} & \textbf{GPAW} &  \textbf{G16}  & \textbf{$\Delta$(\%)} \\
\hline

$\alpha_{x,x}$ & 24.172 & 24.094 & 0.324 & 24.439 & 24.382 & 0.234 \\ 
$\alpha_{y,y}$ & 34.157 & 33.689 & 1.389 & 34.574 & 34.082 & 1.444 \\ 
$A_{y,yy}$ & 10.694 & 10.271 & 4.118 & 11.061 & 10.754 & 2.855 \\ 
$A_{y,yz}$ & -7.797 & -7.636 & 2.108 & -7.735 & -7.479 & 3.423 \\ 
$C_{xx,xx}$ & 101.95 & 100.704 & 1.237 & 104.845 & 103.671 & 1.132 \\ 
$C_{xy,xy}$ & 104.106 & 103.585 & 0.503 & 107.451 & 106.889 & 0.526 \\ 
$C_{yy,yy}$ & 207.858 & 204.958 & 1.415 & 213.457 & 210.499 & 1.405 \\

\hline
\end{tabular}

    \caption{The non-zero components of the irreducible representation of the polarizability tensors. The $\alpha$, $A$ and $C$ tensors have 2, 2, and 3 unique elements respectively for the $C_{3h}$.\cite{buckingham}}
\end{table}
The non-zero components of the reducible tensor representation are as follows\cite{Jahn1937}:
\begin{align*}
    \alpha_{z,z} &= \alpha_{y,y}\\
    A_{y,zz} &= -A_{y,yy}\\
    A_{z,yy} &= A_{y,yz}\\
    A_{z,yz} &= A_{y,zz}\\
    A_{z,zz} &= -A_{z,yy}\\
    C_{xx,yy} &= C_{xx,zz} = \frac{-C_{xx,xx}}{2}\\
    C_{xz,xz} &= C_{xy,xy}\\
    C_{yy,zz} &= -(C_{xx,yy} + C_{yy,yy})\\
    C_{zz,zz} &= C_{yy,yy}\\
    C_{yz,yz} &= \frac{C_{yy,yy}-C_{yy,zz}}{2}\\
\end{align*}

\newpage
\subsection*{$K_{h}-$ $He$}


\begin{minipage}{\linewidth}
    \centering
        \includegraphics[width=0.15\paperwidth]{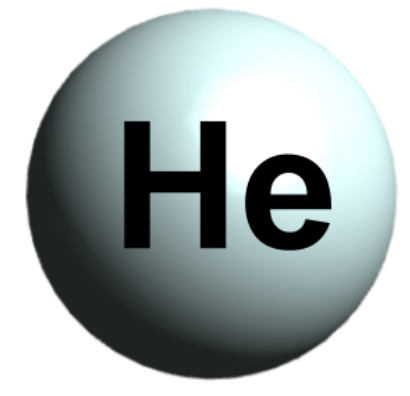}
        \captionof*{figure}{Helium ($He$)}

\end{minipage}

\begin{table}[!ht]
    \centering

    \rowcolors{2}{gray!25}{white}
\renewcommand{\arraystretch}{1.5} 

\begin{tabular}{|l|r|r|r|r|r|r|}
        \rowcolor{blue!30}
\hline
\textbf{Tensor} &  \multicolumn{3}{c|}{\textbf{PBE}}  & \multicolumn{3}{c|}{\textbf{BLYP}}  \\
\cline{2-7}
        \rowcolor{blue!30}

\textbf{Components} &  \textbf{GPAW} &  \textbf{G16} & \textbf{$\Delta$(\%)} & \textbf{GPAW} &  \textbf{G16}  & \textbf{$\Delta$(\%)} \\
\hline
$\alpha_{x,x}$ & 1.58 & 1.572 & 0.509 & 1.572 & 1.562 & 0.64 \\ 
$C_{xx,xx}$ & 1.126 & 0.729 & 54.458 & 1.133 & 0.729 & 55.418 \\

\hline
\end{tabular}

    \caption{The non-zero components of the irreducible representation of the polarizability tensors for Helium, where the He atoms is kept at the origin of the coordinate system. The $\alpha$, $A$ and $C$ tensors have 1, 0, and 1 unique elements respectively for the $K_{h}$.}
\end{table}

The non-zero components of the reducible tensor representation are as follows\cite{buckingham}:
\begin{align*}
    \alpha_{yy} &= \alpha_{zz} = \alpha_{xx}\\
    C_{xx,yy} &= C_{xx,zz} = C_{yy,zz} = \frac{-C_{xx,xx}}{2}\\
    C_{yy,yy} &= C_{zz,zz} = C_{xx,xx} \\
    C_{xy,xy} &= C_{xz,xz} = C_{yz,yz} = \frac{3C_{xx,xx}}{4}\\
\end{align*}

\newpage
\subsection*{$C_{i}-$ $C_{2}H_{2}F_{2}Cl_{2}$}


\begin{minipage}{\linewidth}
    \centering
        \includegraphics[width=0.4\paperwidth]{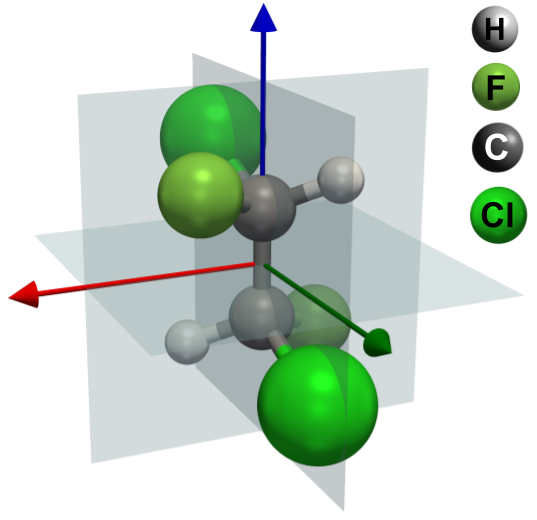}
        \captionof*{figure}{1,2-Dichloro-1,2-difluoroethane ($C_{2}H_{2}F_{2}Cl_{2}$)}

\end{minipage}
\begin{table}[!ht]
    \centering

    \rowcolors{2}{gray!25}{white}
\renewcommand{\arraystretch}{1.5} 

\begin{tabular}{|l|r|r|r|r|r|r|}
        \rowcolor{blue!30}
\hline
\textbf{Tensor} &  \multicolumn{3}{c|}{\textbf{PBE}}  & \multicolumn{3}{c|}{\textbf{BLYP}}  \\
\cline{2-7}
        \rowcolor{blue!30}

\textbf{Components} &  \textbf{GPAW} &  \textbf{G16} & \textbf{$\Delta$(\%)} & \textbf{GPAW} &  \textbf{G16}  & \textbf{$\Delta$(\%)} \\
\hline
$\alpha_{x,x}$ & 51.003 & 50.991 & 0.024 & 51.669 & 51.688 & 0.037 \\ 
$\alpha_{x,y}$ & 2.055 & 2.017 & 1.884 & 2.334 & 2.262 & 3.183 \\ 
$\alpha_{x,z}$ & 0.757 & 0.818 & 7.457 & 1.006 & 0.914 & 10.066 \\ 
$\alpha_{y,y}$ & 73.219 & 71.879 & 1.864 & 74.965 & 73.593 & 1.864 \\ 
$\alpha_{y,z}$ & -12.403 & -12.51 & 0.855 & -12.873 & -13.179 & 2.322 \\ 
$\alpha_{z,z}$ & 61.384 & 60.501 & 1.459 & 62.564 & 61.641 & 1.497 \\ 
$C_{xx,xx}$ & 360.188 & 364.844 & 1.276 & 373.633 & 378.909 & 1.392 \\ 
$C_{xx,xy}$ & 9.165 & 9.088 & 0.847 & 10.388 & 10.471 & 0.793 \\ 
$C_{xx,xz}$ & 24.05 & 23.549 & 2.127 & 26.47 & 25.959 & 1.968 \\ 
$C_{xx,yy}$ & -220.611 & -222.385 & 0.798 & -230.079 & -231.922 & 0.795 \\ 
$C_{xx,yz}$ & 266.882 & 264.793 & 0.789 & 279.129 & 278.786 & 0.123 \\ 
$C_{xy,xy}$ & 359.983 & 359.27 & 0.198 & 374.359 & 373.997 & 0.097 \\ 
$C_{xy,xz}$ & -145.692 & -146.522 & 0.566 & -152.419 & -155.051 & 1.698 \\ 
$C_{xy,yy}$ & -6.767 & -7.209 & 6.131 & -7.923 & -8.242 & 3.87 \\ 
$C_{xy,yz}$ & 22.557 & 22.94 & 1.67 & 25.367 & 24.673 & 2.813 \\ 
$C_{xz,xz}$ & 345.479 & 344.221 & 0.365 & 359.111 & 356.879 & 0.625 \\ 
$C_{xz,yy}$ & -33.312 & -33.059 & 0.765 & -36.845 & -35.244 & 4.543 \\ 
$C_{xz,yz}$ & 32.308 & 32.469 & 0.496 & 36.246 & 35.054 & 3.4 \\ 
$C_{yy,yy}$ & 644.717 & 641.898 & 0.439 & 672.842 & 669.824 & 0.451 \\ 
$C_{yy,yz}$ & -156.271 & -157.024 & 0.48 & -165.029 & -165.101 & 0.044 \\ 
$C_{yz,yz}$ & 604.414 & 602.334 & 0.345 & 631.718 & 628.205 & 0.559 \\

\hline
\end{tabular}

    \caption{The non-zero components of the irreducible representation of the polarizability tensors. The $\alpha$, $A$ and $C$ tensors have 6, 0, and 15 unique elements respectively for the $C_{i}$.\cite{buckingham}}
\end{table}

The non-zero components of the reducible tensor representation are as follows:
\begin{align*}
    C_{xx,zz} &= -(C_{xx,xx} + C_{xx,yy})\\
    C_{xy,zz} &= -(C_{xy,yy} + C_{xx,xy})\\
    C_{xz,zz} &= -(C_{xz,yy} + C_{xx,xz})\\
    C_{yy,zz} &= -(C_{yy,yy} + C_{xx,yy})\\
    C_{yz,zz} &= -(C_{yy,yz} + C_{xx,yz})\\
    C_{zz,zz} &= -(C_{xx,zz} + C_{yy,zz})\\
\end{align*}

\clearpage

\newpage
\subsection*{$O_{h}-$ $C_{8}H_{8}$}


\begin{minipage}{\linewidth}
    \centering
        \includegraphics[width=0.4\paperwidth]{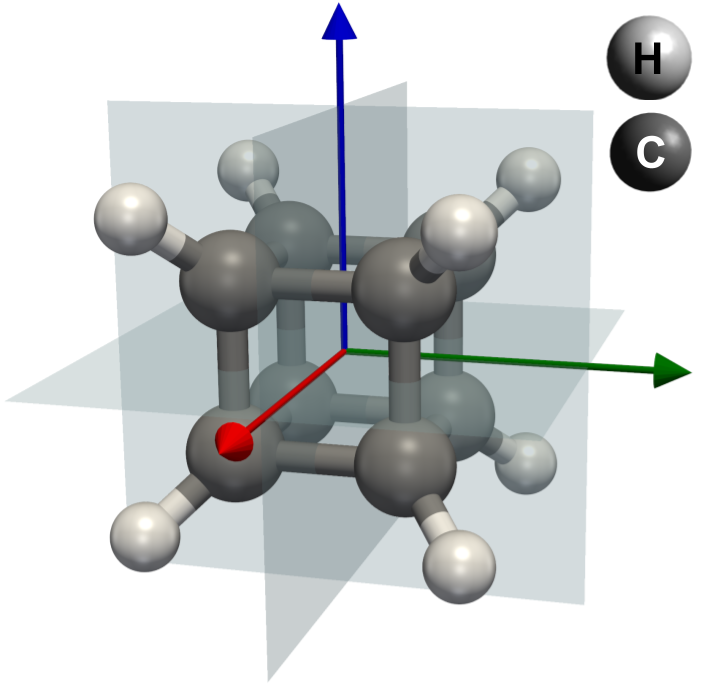}
        \captionof*{figure}{Cubane ($C_{8}H_{8}$)}

\end{minipage}
\begin{table}[!ht]
    \centering

    \rowcolors{2}{gray!25}{white}
\renewcommand{\arraystretch}{1.5} 

\begin{tabular}{|l|r|r|r|r|r|r|}
        \rowcolor{blue!30}
\hline
\textbf{Tensor} &  \multicolumn{3}{c|}{\textbf{PBE}}  & \multicolumn{3}{c|}{\textbf{BLYP}}  \\
\cline{2-7}
        \rowcolor{blue!30}

\textbf{Components} &  \textbf{GPAW} &  \textbf{G16} & \textbf{$\Delta$(\%)} & \textbf{GPAW} &  \textbf{G16}  & \textbf{$\Delta$(\%)} \\
\hline
$\alpha_{x,x}$ & 81.113 & 80.586 & 0.654 & 82.057 & 81.58 & 0.585 \\ 
$C_{xx,xx}$ & 444.448 & 448.782 & 0.966 & 452.773 & 457.729 & 1.083 \\ 
$C_{xy,xy}$ & 658.607 & 659.793 & 0.18 & 673.047 & 674.901 & 0.275 \\

\hline
\end{tabular}

    \caption{The non-zero components of the irreducible representation of the polarizability tensors. The $\alpha$, $A$ and $C$ tensors have 1, 0, and 2 unique elements respectively for the $O_{h}$\cite{buckingham}.}
\end{table}

The non-zero components of the reducible tensor representation are as follows\cite{Jahn1937}:
\begin{align*}
    \alpha_{yy} &= \alpha_{zz} = \alpha_{xx}\\
    C_{yy,yy} &= C_{zz,zz} = C_{xx,xx} \\
    C_{xx,yy} &= C_{xx,zz} = \frac{-C_{xx,xx}}{2}\\
    C_{yy,zz} &= C_{xx,yy}\\
    C_{yz,yz} &= C_{xz,xz} = C_{xy,xy}\\
\end{align*}

\newpage
\subsection*{$T_{d}-$ $CH_{4}$}


\begin{minipage}{\linewidth}
    \centering
        \includegraphics[width=0.4\paperwidth]{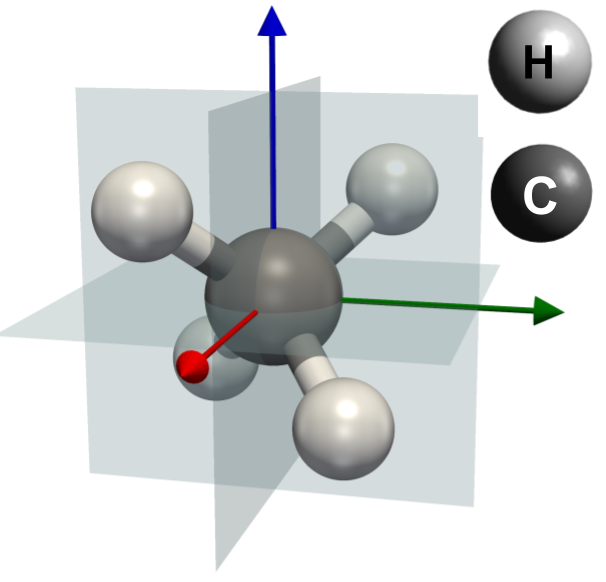}
        \captionof*{figure}{Methane ($CH_{4}$)}

\end{minipage}
\begin{table}[!ht]
    \centering

    \rowcolors{2}{gray!25}{white}
\renewcommand{\arraystretch}{1.5} 

\begin{tabular}{|l|r|r|r|r|r|r|}
        \rowcolor{blue!30}
\hline
\textbf{Tensor} &  \multicolumn{3}{c|}{\textbf{PBE}}  & \multicolumn{3}{c|}{\textbf{BLYP}}  \\
\cline{2-7}
        \rowcolor{blue!30}

\textbf{Components} &  \textbf{GPAW} &  \textbf{G16} & \textbf{$\Delta$(\%)} & \textbf{GPAW} &  \textbf{G16}  & \textbf{$\Delta$(\%)} \\
\hline
$\alpha_{x,x}$ & 17.768 & 17.698 & 0.396 & 17.788 & 17.721 & 0.378 \\ 
$A_{x,yz}$ & -9.771 & -9.817 & 0.469 & -10.076 & -10.136 & 0.592 \\ 
$C_{xx,xx}$ & 42.763 & 42.698 & 0.152 & 43.466 & 43.405 & 0.141 \\ 
$C_{xy,xy}$ & 37.122 & 37.042 & 0.216 & 37.925 & 37.832 & 0.246 \\

\hline
\end{tabular}

    \caption{The non-zero components of the irreducible representation of the polarizability tensors. The $\alpha$, $A$ and $C$ tensors have 1, 1, and 2 unique elements respectively for the $T_{d}$\cite{buckingham}.}
\end{table}

The non-zero components of the reducible tensor representation are as follows\cite{Jahn1937}:
\begin{align*}
    \alpha_{y,y} &= \alpha_{z,z} = \alpha_{x,x}\\
    A_{z,xy} &= A_{y,xz} = A_{x,yz}\\
    C_{yy,yy} &= C_{zz,zz} = C_{xx,xx} \\
    C_{xx,yy} &= C_{xx,zz} = \frac{-C_{xx,xx}}{2}\\
    C_{yy,zz} &= C_{xx,yy}\\
    C_{yz,yz} &= C_{xz,xz} = C_{xy,xy}\\
\end{align*}

\bibliography{supplement}